\documentclass[12pt, centerh1]{article}

\RequirePackage[colorlinks,citecolor=blue,urlcolor=blue]{hyperref}
\usepackage{amsmath, amssymb,natbib}
\usepackage{subcaption}
\usepackage{graphicx,bm}
\usepackage{color}
\usepackage{subcaption}
 \usepackage[table]{xcolor}
\usepackage{longtable}
\usepackage{amsthm}
\usepackage[mathscr]{euscript}
\usepackage{relsize}
\newcolumntype{P}[1]{>{\centering\arraybackslash}p{#1}}
\usepackage{rotating}
\usepackage{eurosym}
\usepackage{colonequals}
\usepackage{bbm}
\usepackage{pbox}
\usepackage{booktabs}
\usepackage{dsfont}
\usepackage{authblk}

\DeclareMathOperator*{\argmax}{argmax}

\newcommand{\bmf}{\bm{f}} 

\title{ Defying the Circadian Rhythm: Clustering Participant Telemetry in the UK Biobank Data} %

\author[1]{Nikola Po\v cu\v ca}
\author[2]{Mark Farrell}
\author[1]{Paul D. McNicholas}

\affil[1]{Department of Mathematics and Statistics, McMaster University, Ontario, Canada}
\affil[2]{Queen's Management School, Queen's University, Belfast, Northern Ireland}

\newcommand{\hbb}{\bm{H}}

\newcommand{\MM}{\bm{M}}
\newcommand{\MV}{\bm{V}}
\newcommand{\MU}{\bm{U}}

\begin{document}

\maketitle
\begin{abstract}
The UK Biobank dataset follows over $500,000$ volunteers and contains a diverse set of information related to societal outcomes. Among this vast collection, a large quantity of telemetry collected from wrist-worn accelerometers provides a snapshot of participant activity. Using this data, a population of shift workers, subjected to disrupted circadian rhythms, is analysed using a mixture model-based approach to yield protective effects from physical activity on survival outcomes. In this paper, we develop a scalable, standardized, and unique methodology that efficiently clusters a vast quantity of participant telemetry. By building upon the work of \cite{doherty2017}, we introduce a standardized, low-dimensional feature for clustering purposes. Participants are clustered using a matrix variate mixture model-based approach. Once clustered, survival analysis is performed to demonstrate distinct lifetime outcomes for individuals within each cluster. In summary, we process, cluster, and analyse a subset of UK Biobank participants to show the protective effects from physical activity on circadian disrupted individuals.

\noindent\textbf{Keywords}: UK Biobank, accelerometer data, matrix variate, telemetric data, Cox regression, hazard ratios, big data, high-dimensional data, large scale data processing.
\end{abstract}

\section{Introduction}
The UK BioBank dataset is a large collection of participant data with a broad selection of information that is
relevant for statistical investigation \citep{biobank2014}.
The use of accelerometer data for predicting lifetime outcomes has become a key research focus in the last decade.
 There is an overwhelming consensus that physical inactivity has detrimental effects on lifetime outcomes \citep{lee2012}. Those include inter alia; breast cancer, coronary heart disease, and type 2 diabetes \citep{la2012}. As a result, physical inactivity directly affects life expectancy \citep{cunningham2020}; leading to early death and onset of disease.
 
Additional factors, often not considered, are circadian rhythm defying lifestyle choices. Many workers who undertake their job in some form of shift work or extended hours have implications on respective sleeping patterns. Studies have shown that these lifestyles can have negative health effects \citep{harrington2001,dembe2005,fan2020}. However, \cite{roveda2017} showed a
protective effect where physical activity may be beneficial against the detrimental health effects typically associated with sleep disruption. Such studies are not uncommon as found in \cite{yamanaka2006}, \cite{montaruli2017} and \cite{nohara2015}.

As highlighted by \cite{doherty2017}, a major short coming of studies in this area have been based on self-reported evidence. As a consequence, such studies are subject to inaccuracies as self-reported evidence is heavily biased; particularly when concerning physical activity as outlined in \cite{brenner2016}. Again, \cite{doherty2017} emphasizes the growing use of objective measures for physical activity in the form of telemetry collected from wrist-worn accelerometers. However, the use of accelerometer telemetry beckons another issue. The sheer overwhelming quantity and vastness of telemetric data. At the time of analysis, the size of accelerometer data totalled to 26 terabytes of memory. Each participant generates approximately 217 megabytes of data or, the equivalent of two and a half hours of social media consumption \citep{index2016global}. For specifics regarding wear time compliance, \cite{doherty2017} provides a thorough analysis  across age and sex. 

 From a computational perspective, the size of data poses massive challenges to process, manage, and analyse lifetime outcomes. Our paper builds upon \cite{doherty2017} and \cite{willetts2018}, by developing a standardized, low-dimensional feature that captures physical activity behaviour from accelerometer telemetry. We use a mixture model-based approach to cluster and identify participant sub-groups. Finally, we perform survival analysis to demonstrate how physical inactivity directly affects lifetime outcomes. 

\section{Methodology}
This section is broken down in chronological order on how the analysis is performed. Firstly, a cohort is selected where lifetime outcomes are well known from an occupational health perspective. Second, detail is provided on the use of the activity extraction tool outlined in \cite{doherty2017}. Thirdly, a low-dimensional feature that standardizes and captures physical activity is developed.  Furthermore, a  mixture model-based approach is used to cluster said feature. Finally, we summarize two popularized methods used for survival analysis.

\subsection{Study Population Selection}
There is a well known phenomenon that life expectancy is lower among workers that defy their natural circadian rhythm through lifestyle factors such as shift work or extended hours \citep{gu2015}. Consequently this results in unusual sleeping patterns. Several studies from an occupational health perspective, highlight cases of early onset non-communicable diseases, thus lowering their life expectancy \citep{ijaz2013}.
Within the UK Biobank dataset, a large subset of participants reported their employment history regarding shift work. 
We investigate this phenomenon within the aformentioned population to achieve a consesus with previous literature.

At the time of analysis the UK Biobank dataset contained $502,536$ individuals. 
First, a subset of participants is selected based on their response to a questionnaire regarding employment history. 
The subset is split between two groups, regular shift, and late shift. Groups are defined by the respective participant response to a questionnaire. We define the regular shift group as participants who's response was having never or rarely experienced night shift work. In contrast, we define night shift workers as participants who's response was having always experienced night shift work. Second, individuals are isolated who have had their accelerometer telemetry collected.
Finally, all individuals who meet both of the aforementioned data requirements are grouped; yielding $5,507$ participants. 
This population of participants is defined as SW-0, pertaining to the raw, unclustered shift worker participants.
Of these participants, $4,471$ are regular shift workers, and $1,036$ are late. 
We draw attention to the unbalanced nature of the two groups where regular shift workers outnumber late shift approximately 4 to 1. 
Nevertheless, these sample sizes are well within a valid range suitable for survival analysis.

 \begin{table}
\parbox{.45\linewidth}{
\centering
\caption{Biological sex counts for regular shift workers.}
\label{table:sex-reg}
\begin{tabular}{|c|c|c|}
\hline
Sex & Female &  Male \\
\hline
Counts & $2505$  & $1966$ \\
\% & $56.02$ & $43.98$  \\ 
\hline
\end{tabular}
}
\hfill
\parbox{.45\linewidth}{
\centering
\caption{Biological sex counts for late shift workers.}
\label{table:sex-late}
\begin{tabular}{|c|c|c|}
\hline
Sex & Female &  Male \\
\hline
Counts & $419$  &  $617$  \\
\% & $40.45$ & $59.55$  \\ 
\hline
\end{tabular}
}
\end{table}
Tables \ref{table:sex-reg} and \ref{table:sex-late} show a fairly even split between regular and late shift workers. There is a clear majority of males among late shift workers. As a consequence, known survival outcomes based on sex differences may play a role in our analysis. Typically, literature has shown to favour females over males \citep{women2020,antero2020female,di2020gender}. These differences are taken into account during our survival analysis. However, as a simplification, we proceed without separating based on sex during our clustering procedures.
\newpage

\subsection{Data Processing} \label{sec:2.2}
Recent developments on extracting behavioural patterns from the UK BioBank dataset has yielded an opportunity for 
prediction of lifetime outcomes. \cite{doherty2017} provides a powerful tool that allows for a participant's telemetry to be 
processed into a more manageable form. This tool classifies each $5$ second interval into one of the following activities:
sleep, moderate, light tasks, sedentary, or walking. The resultant output of the tool contains the following. 
A csv file of accelerometer data containing a column of milli-gravity units, and, the classified activity as a category for every time interval.
For each individual participant within SW-0, their respective cwa files are isolated totalling 1.2 terabytes in memory.
Using the behavioural extraction tool, all $5,507$ cwa files are processed into their respective csv files. 
Issues arose during the process, yielding only $5,029$ csv files. 
These individuals are defined as SW-P, pertaining to their corresponding processed csv files. 
There is an overall $8.35 \% $ drop in individuals for SW-P when compared to SW-0. These issues are not uncommon and coincide with the previous results of \cite{doherty2017}, where $6.7 \%$ of individuals had insufficient wear time. For further specifics on wear time compliance, age, sex and other participant characteristics, see \cite{doherty2017}. 

\subsection{Feature Development}
Each processed csv file contains force measurements for each five second interval. \cite{doherty2017} points out that each csv file does not have a standard number of five second intervals. Different participants wore their accelerometers for different periods of time. To rectify this issue, we develop a feature known as a probability force heat map (force map). A force map is a two dimensional kernel density estimation (KDE) of all force measurements, and, their respective derivative \citep{kde2019}. This type of feature has been well studied from an engineering and actuarial perspective \citep{kamble2009,mario2017}. Our force map is constructed as follows.  Let $f_t$ denote force measurement at time interval $t$. Force is measured in milli-gravity units (mg), where it is averaged over a $5$ second time interval. Let $f^{'}_{t}$ denote the numerical first derivative calculated as 
$$ \quad f^{'}_{t} = \frac{f_t - f_{t-1}}{5}.$$ 
Note the denominator, as each interval differs by $5$ seconds. $f^{'}_{t}$ is considered to be average change in force over some time interval $t$. \cite{mario2017} elaborates that this calculation is determined by the ``average" change over the time interval, and not reminiscent of the true instantaneous measurement. To construct a force map,  let $R$ be some rectangle where $\bm{f}_t = ({f_t , f^{'}_{t}} )$ is considered to be values of a two dimensional coordinate system. Next, construct a partitioning of $R$ into $M$ equally sized rectangles as 
$$R = \bigcup_{m=1}^M R_m ,\quad R_m \cap R_{m'} = \emptyset, \quad \forall m \neq m'.$$
Finally, consider a probability distribution of $F \in \mathcal{P}(R)$ having probability weight 
$$x_m = \int_{R_m} d F  \geq 0, \quad m = 1, \dots, M , \quad \text{satisfying } \sum_{m=1}^M x_m = 1.  $$ 
The selection of $F$ can allow for smooth approximations to the natural behaviour. As \cite{pocuca2019} demonstrates, the selection of a two dimensional Gaussian kernel is superior to other methods used in \cite{mario2017} and \cite{gao2019}. 

Let $\bmf_i = \{ \bmf_{it} \}_{t=1}^T $, corresponding to the bivariate collection of all force measurements for a particular participant $i$. Next, consider the use of KDE for the problem of estimating an unknown joint probability density $p( \bmf )$ on the space $R \subseteq \mathbb{R}^2$. Let $K$ be a bivariate function defined on the space $\mathbb{R}^2$. Furthermore, let $\bm{H}$ be a constant, positive definite, symmetric matrix defined as the smoothness parameter. The KDE of $p( \bmf  )$ is written as 
$$ \hat{p}_{\bm{H}}(\bmf ) = \frac{1}{T}\sum_{t=1}^{T} |\bm{H}|^{-\frac{1}{2}} K\left(\bm{H}^{-\frac{1}{2}}(\bmf  -\bmf_{it})\right),$$ $\bmf \in R$.  
Due to popularity and radial symmetrical properties, the standard bivariate Gaussian distribution is selected as the kernel function $K$. As a result, $K$ is formulated as 
$$ K \left(\bm{H}^{-\frac{1}{2}}(\bmf  -\bmf_{it})\right) =  \left(2 \pi \right)^{-1}  \text{exp} \left\{ -\frac{1}{2} (\bmf  -\bmf_{it})^{\top}  \bm{H}^{-1} (\bmf  - \bmf_{it})  \right\}.  $$
This selection of $K$ allows the kernel estimator to be the weighted sum of normal densities centred at force measurements $\bmf_{it}$. In summary, the true density $p( \bmf  )$ is approximated using the  kernel estimator 
$$ \hat{p}_{\bm{H}}(\bmf) = \frac{1}{T} \sum_{t=1}^{T} \phi\left(\bmf; \bmf_{it}, \bm{H} \right).$$

Here, $\phi$ is the Gaussian density with covariance matrix $\hbb$. The selection of the smoothness matrix $\hbb$ affects both the shape and orientation of the kernels on the two dimensional space.  The bandwidth is a key parameter for optimizing performance of KDE. For the purposes of simplicity, the normal scale selector \citep{normscale} is selected as
$$\hbb_{NS} = \left( \frac{1}{T} \right)^{\frac{1}{3}} \hat{\bm{\Sigma}},$$
where $\hat{\bm{\Sigma}}$ is considered to be the sample covariance matrix. The use of this method for estimating the probability density function is superior to that of a histogram or an empirical kernel. Furthermore, the KDE method captures the continuous nature of data even when latency is large. For specifics, see \cite{pocuca2019}. For implementation, see the \texttt{MASS} package regarding the use of two dimensional KDE \citep{mass}.  

Finally, construct the matrix variate object $\bm{X}_i$ whose matrix entries directly correspond to the probability weights $x_{im}$ over region $R_m$. Probability weights are calculated as
$$x_{im} = \int_{R_m}   \hat{p}_{\bm{H}_{NS}}(\bmf)  \partial \bmf   > 0, \quad m = 1, \dots, M , \quad \text{satisfying } \sum_{m=1}^M x_{im} = 1, \: \text{where }\bm{X}_i = \{x_{im} \}_{m=1}^M. $$

This matrix variate object is defined as the force map for participant $i$. Note that this extension differs from the original construction by \cite{mario2017} in one key distinction. That distinction being the use of KDE to estimate the joint density on $R$. The choice of a smooth kernel allows the matrix $\bm{X}_i$  to have non-zero entries which more accurately resemble a participant's behaviour.

\begin{figure}[!htb]

  \begin{subfigure}[t]{0.45\textwidth}
        \centering
        \caption{Force data}
		\includegraphics[scale = 0.37]{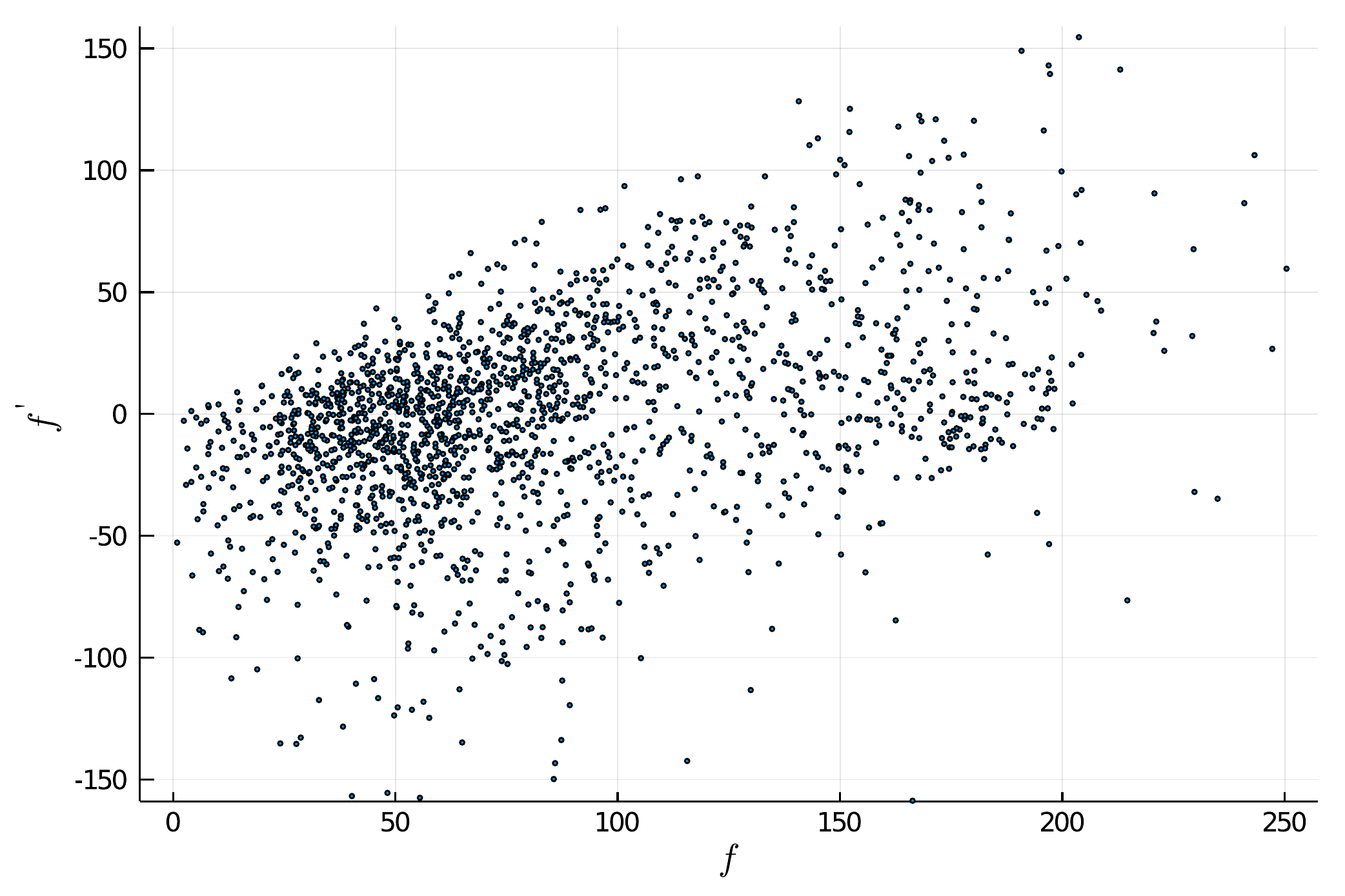}
        \label{fig:walk_pre}
    \end{subfigure} %
        ~ 
                   \begin{subfigure}[t]{0.45\textwidth}
        \centering
        \caption{Force map}
		\includegraphics[scale = 0.37]{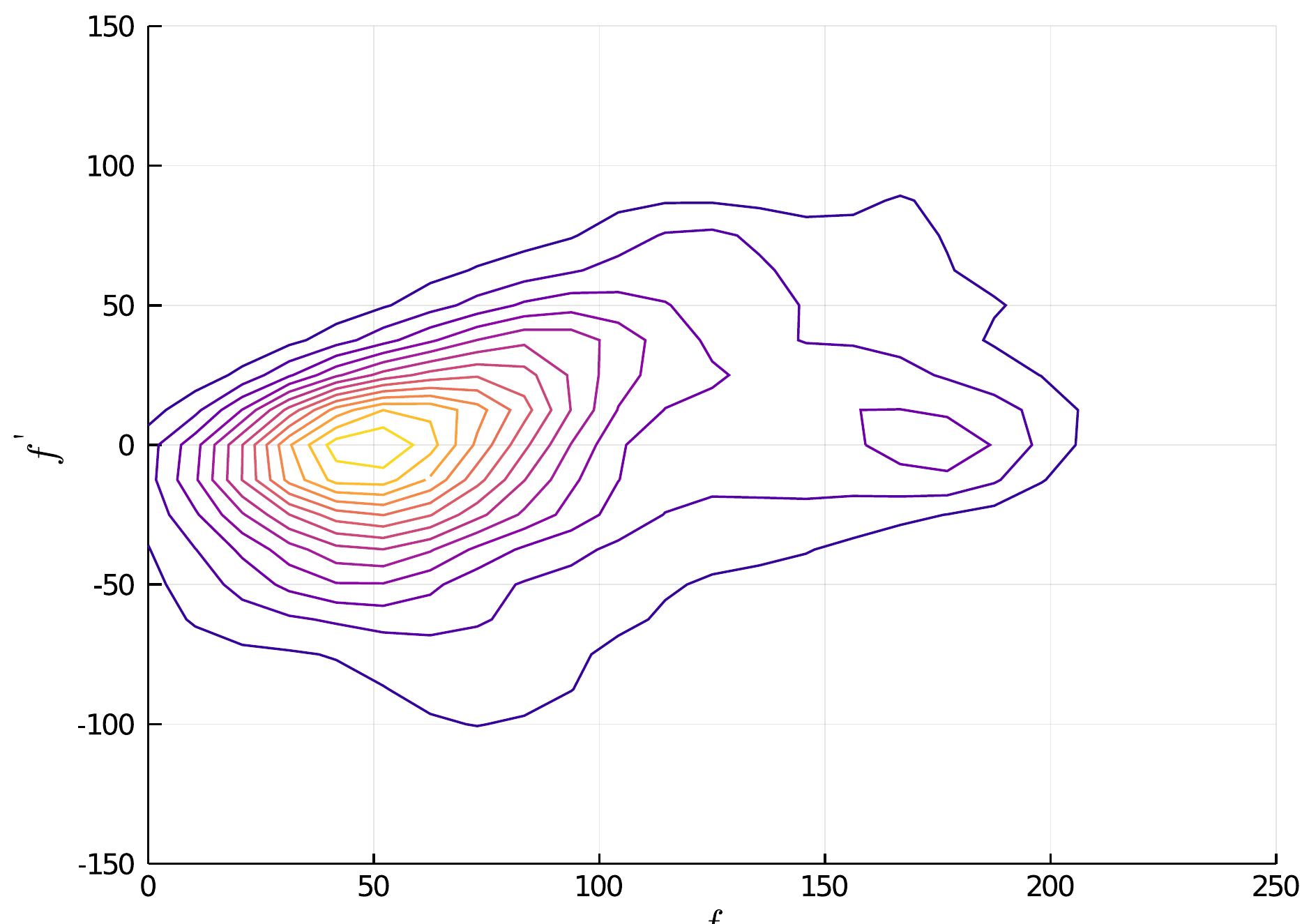}
        \label{fig:walk_after}
    \end{subfigure}%

\caption{Walking force map construction for participant $60004621$.}
\label{fig:forcemap_construction}
\end{figure}

\begin{figure}[!htb]
	\centering
  \begin{subfigure}[t]{0.45\textwidth}
        \centering
        \caption{Sleep}
		\includegraphics[scale = 0.37]{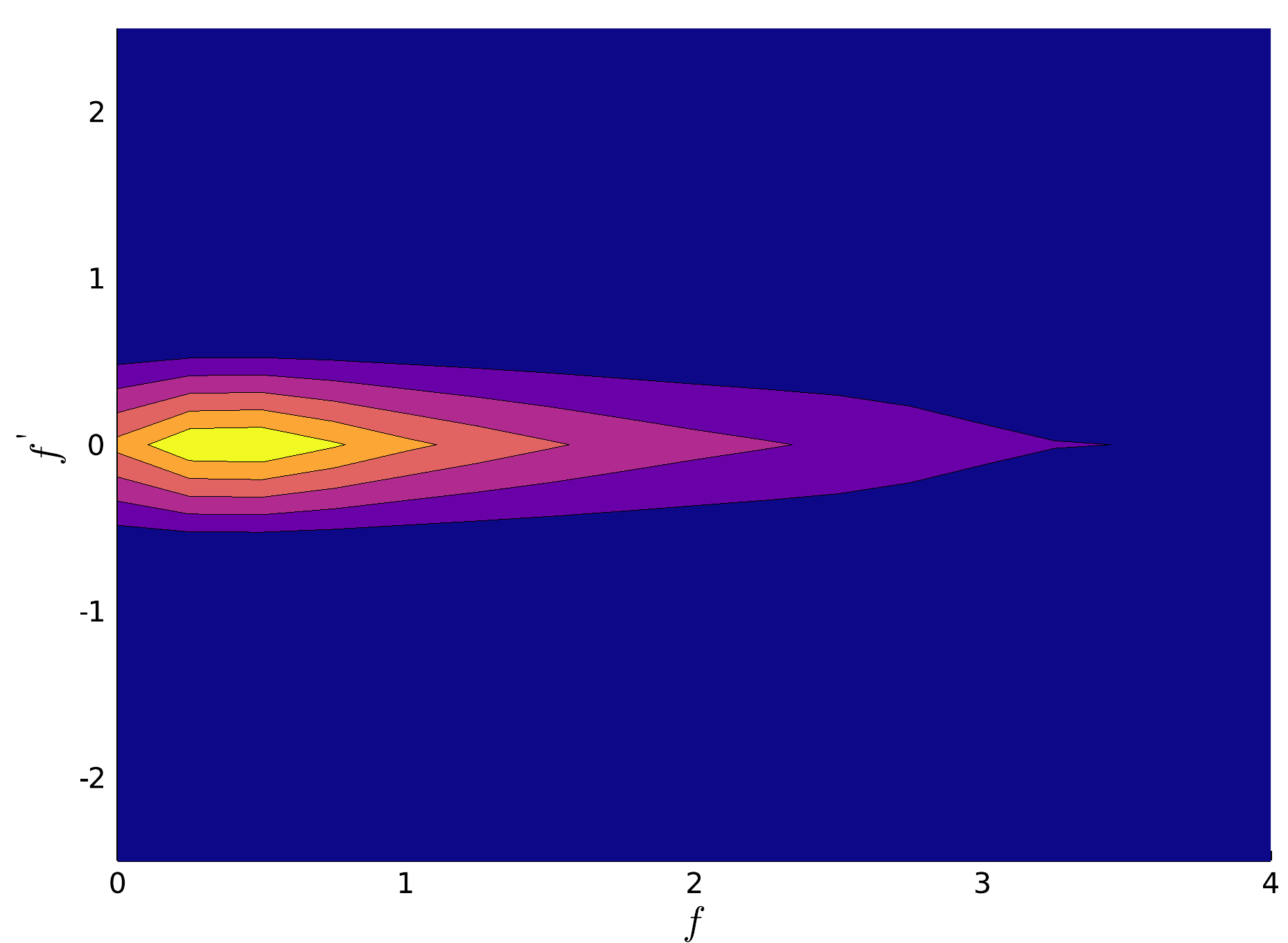}
        \label{fig:sleep}
    \end{subfigure} %
    \hspace{-0.3cm}
        \begin{subfigure}[t]{0.45\textwidth}
        \centering
        \caption{Sedentary}
		\includegraphics[scale = 0.37]{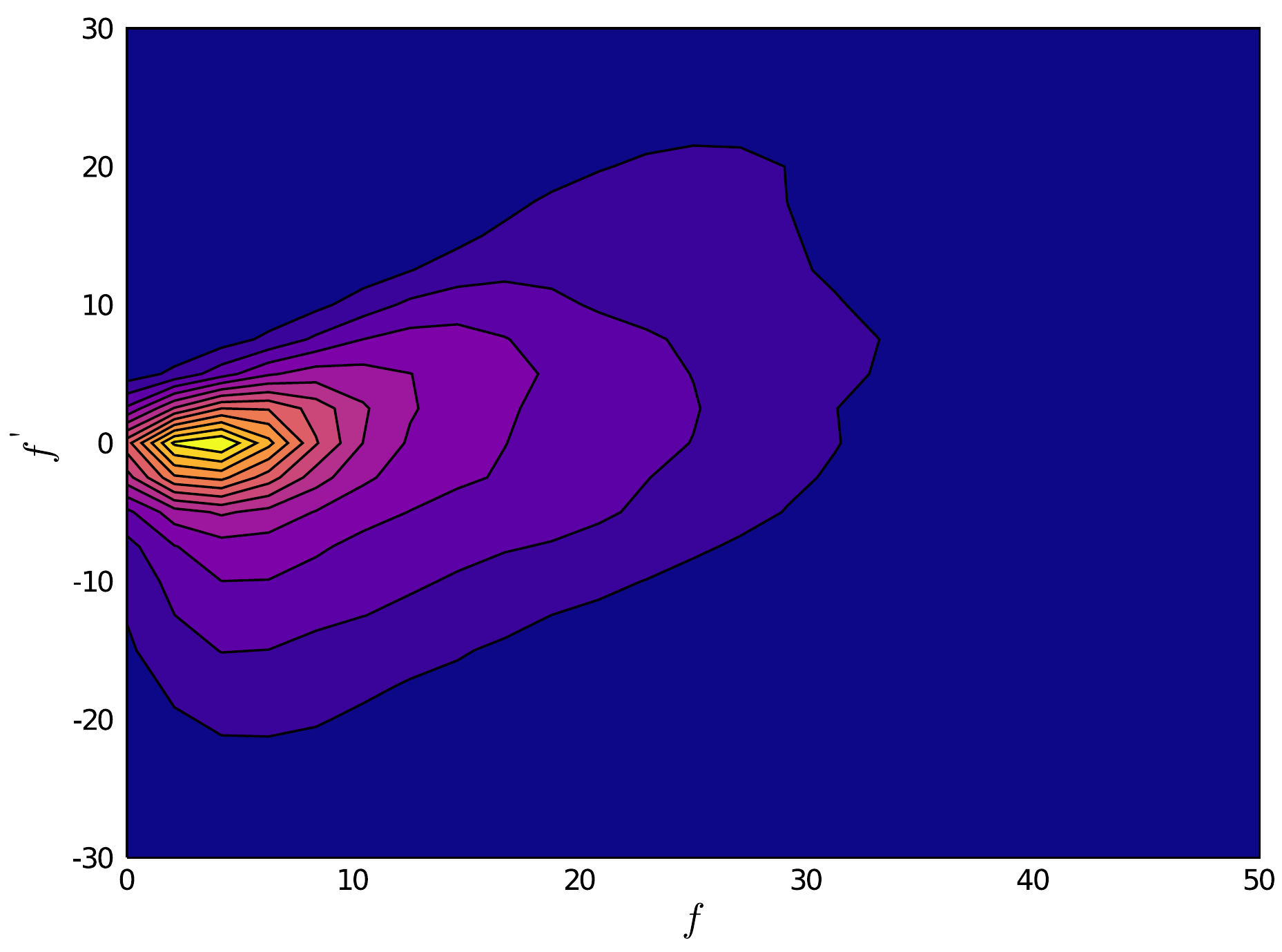}
        \label{fig:sed}
    \end{subfigure}%
    \\
           \begin{subfigure}[t]{0.45\textwidth}
        \centering
        \caption{Light Tasks}
		\includegraphics[scale = 0.37]{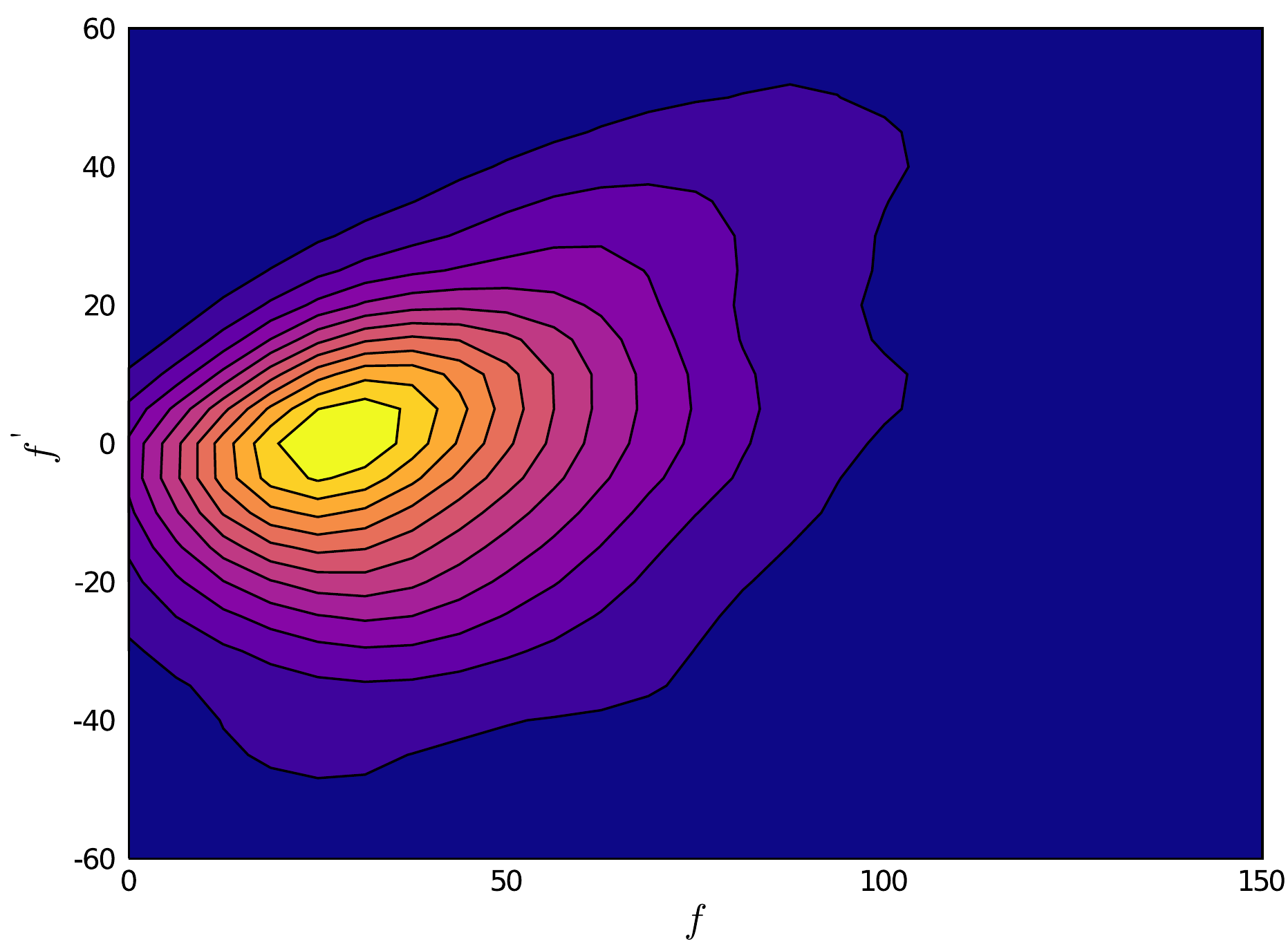}
        \label{fig:light}
    \end{subfigure}%
               \begin{subfigure}[t]{0.45\textwidth}
        \centering
        \caption{Moderate Tasks}
		\includegraphics[scale = 0.37]{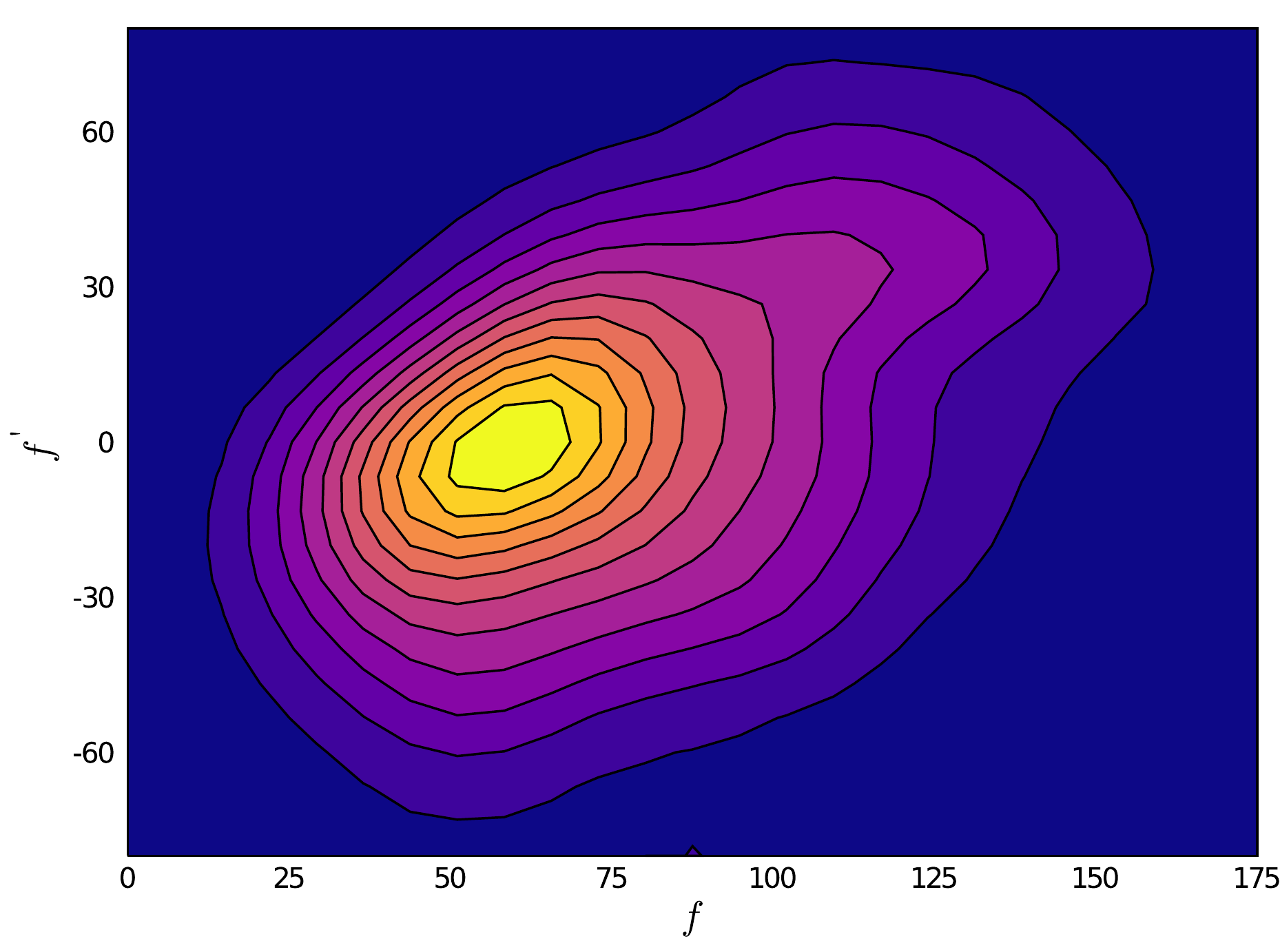}
        \label{fig:mod}
    \end{subfigure}%

\caption{Force map contours for participant $60004621$ by predicted behaviour.}
\label{fig:forcemaps}
\end{figure}

Figure \ref{fig:forcemap_construction} visualizes the force map construction for walking data. For this specific participant, their walking behaviour in Figure \ref{fig:walk_pre} exhibits forces mostly at 50 units, while their change in force is roughly between -20 to 20 units. As a result, their walking force map in Figure \ref{fig:walk_after} has a higher probability measure at this location. In addition, they also appear to have another localized area of force measurements centred at 175 units. This is again reflected in the force map in Figure \ref{fig:walk_after}. In summary, for this participant, we capture their walking behaviour characterized by this force map. Repeating the same process for each behaviour, we visualize force maps as shown in Figure \ref{fig:forcemaps}.

We draw attention to the scale for each plot, as sleep, and sedentary maps have smaller force readings. In contrast, light and moderate tasks have higher force readings. For this participant, we demonstrate that force maps capture the desired behaviour, as sleep is more concentrated, while light to moderate tasks have a more spread estimated probability distribution. As desired, the force map reflects the natural behaviour of the participant. As a consequence of using this approach, all participant data is standardized into a $25$ by $25$ matrix suitable for clustering. Essentially the force maps, constructed from accelerometer data, clearly illustrate each of the 5 aforementioned participant activities. Furthermore, these force maps are visually interpretable, characterizing a unique exhibit of participant behaviour.

\newpage
\subsection{Model-based Clustering}

The framework of model-based clustering utilizes the underlying assumption that a finite mixture model embodies the representation of heterogeneous data. Consider a random variable $\bm{\mathcal{X}}$ from a $G$-component finite mixture model with probability density function of the form 
\begin{equation} p(\bm{x} | \bm{\vartheta}) = \sum_{g = 1}^G \pi_g p_g(\bm{x} | \bm{\theta_g}), \label{mixture} \end{equation}
where $ \bm{\vartheta} = \{\pi_1, \dots, \pi_G, \bm{\theta}_1, \dots, \bm{\theta}_G\}$,  $\bm{x}$ is a realization of $\bm{\mathcal{X}}$, $\pi_g$ is a mixing proportion where $\pi_g > 0$, $\sum_{g=1}^G\pi_g = 1$, and  $p_g$ is a probability density function parametrized by $\bm{\theta}_g$. The distribution for each cluster is usually taken to be the same density in   \eqref{mixture} and is simplified as $p_g(\bm{x} | \bm{\theta}_g) = p(\bm{x} | \bm{\theta}_g) \quad \forall g$. 

For the purposes of clustering high dimensional data, issues arise due to the curse of dimensionality \citep{curseDim}. A standard approach is to reduce the number of dimensions by considering a series of underlying factors with a lesser dimension \citep{spearman}. Let $\bm{\mathcal{X}}_i$ represent an $r$ dimensional random vector, with $\bm{x}_i$ as a realization. The factor analyzers model for $\bm{\mathcal{X}}_1$, \dots , $\bm{\mathcal{X}}_N$, is given by 
$$\bm{\mathcal{X}}_i =  \bm{\mu} + \bm{\Lambda} \bm{U}_i + \bm{\epsilon}_i,$$ 
where $\bm{\mu}$ is a mean location vector, $ \bm{\Lambda}$ is a  $r \times s$ matrix of factor loadings, with $s < r$, $\bm{U}_i \sim \mathcal{N}_r( \bm{0},\bm{I})$ denoting the latent factors, and $\bm{\epsilon}_i \sim \mathcal{N}_r(\bm{0},\bm{\Psi}) $  where $\bm{\Psi}= \text{diag}(\psi_1, \psi_2, . . . , \psi_r)$. Here,  $ \mathcal{N}_r$ denotes the $r$-dimensional multivariate normal. Furthermore, the latent factors $\bm{U}_i$ and noise $\bm{\epsilon}_i$ are independent of each other. It is noted that the probabilistic principal component analysis (PPCA) is a special case of the factor analysis model with a specific isotropic constraint on $\bm{\Psi}$ \citep{ppcaFact}.  The factor analyzers model is a flexible extension of PPCA. By removing the isotropic constraint on $\bm{\Psi}$, this allows parameters to vary. The factor analyzers model is considered to be the best choice for dealing with telemetric data as it is highly efficient in reducing dimensionality \citep{ppcaRef}.

Suppose a matrix is considered to be an observation sampled from a distribution. Naturally, an appropriately sized matrix variate distribution should be considered to model randomness. Consider the matrix variate normal distribution \citep{gupta}. Let $\mathscr{X}$ be a random variable with an $r\times c$ matrix $\bm{X}$ as a realization. As a result, $\mathscr{X}$ is distributed according to a matrix variate distribution. 
The random matrix $\bm{X}(r \times c)$ is said to have a matrix variate normal distribution with mean matrix $\MM  (r \times c)$ and covariance matrix $ \bm{\Psi} \otimes \bm{\Sigma}$. Each matrix is appropriately sized as $\bm{\Sigma} (r \times r)$, $\bm{\Psi} (c \times c)$, where $\text{vec}(\mathscr{X}) \sim \mathcal{N}_{rc}(\text{vec}(\MM'),\bm{ \Psi}  \otimes \bm{ \Sigma} ) $. Here, $\otimes$ refers to the Kronecker Product and $\text{vec}$ denotes the vectorization of a matrix. Given this specification, the density is formulated as  
$$ \varphi(\bm{X}; \MM, \bm{\Psi} \otimes \bm{\Sigma}) = \frac{1}{(2\pi)^{ \frac{rc}{2} } |\bm{\Psi}|^{\frac{r}{2}} |\bm{\Sigma}|^{\frac{c}{2}}}\exp \bigg\{  - \frac{ 1 }{ 2 } \text{tr} \big(  \bm{\Psi}^{ -1 } (\bm{X} - \MM)^{'} \bm{\Sigma}^{-1} (\bm{X}-\MM) \big) \bigg\}. $$
The matrix variate normal distribution is defined through a vectorization of a multivariate normal. Note that the covariance matrices of row and column are non-unique as they are defined through a Kronecker product \citep{dutMLE}. As a result, both densities are parametrized by the product and not individual co-variance matrices \citep{gupta}. There are benefits for using a matrix variate representation. The main benefit of the matrix variate approach is the speed at which model parameters are estimated in high dimensional settings.  

Due to the issues of high dimensionality, an analogous extension of the factor analyzers model for matrix variate data is implemented. The mixtures of matrix variate bilinear factor analyzers model (MBI) is a powerful approach for dealing with both high-dimensional data, and the presence of a mixture of populations \citep{bifact}. 
Suppose latent factors of size $s < r$, $v < c$, for a matrix variate random variable constitute the data with probability $\pi_g$ of occurring as 
 $$ \bm{X}_i = \MM_g + \bm{A}_g \bm{W}_{ig}\bm{B}'_g + \bm{A}_g \mathcal{E}_{ig}^{B} +\mathcal{E}_{ig}^{A} \bm{B}_g' + \mathcal{E}_{ig}, $$
where $\MM_g (r,c) $ is the mean matrix, $\bm{W}_{ig} (s,v) \sim \mathcal{N}_{s \times v}(\bm{0}, \bm{I}_s,\bm{ I}_v) $ is a matrix random variate of latent factors,  $\bm{A}_g (r \times s)$ are column factor loadings, and $\bm{B}_g (c \times v)$ are row factor loadings respectively. Finally, the noise is distributed according to 
 $$\mathcal{E}_{ig}^A \sim \mathcal{N}_{r \times v}(\bm{0}, \MU_g,\bm{I}_v),\quad
 \mathcal{E}_{ig}^B \sim \mathcal{N}_{s \times c}(\bm{0}, \bm{I}_s,\MV_g),\quad
 \mathcal{E}_{ig} \sim \mathcal{N}_{r \times c}(\bm{0}, \MU_g,\MV_g). $$ 
For applications in accelerometer telemetry, the MBI model is used to cluster matrix variate objects pertaining to the heterogeneous population of participants. 

The estimation procedure for MBI is based on local maximum likelihood estimation. The most common approach for estimating finite mixture models is with the expectation maximization (EM) algorithm \citep{emAlgo}. However, when dealing with latent factor models, \cite{pars} uses the alternating expectation–conditional maximization algorithm \citep[AECM;][]{aecm}. Estimation of parameters pertaining to the MBI model is performed as follows.  Consider a latent variable $Z_{ig}$ denoting membership of observation $\bm{X}_i$ belonging to group $g$ as, 
$$ Z_{ig} = \begin{cases}
      \quad  1, &  \text{observation $\bm{X}_i$ belongs to group $g$}\\
     \quad 0 ,  & \text{otherwise}.
   \end{cases}$$
For example, the component membership for observation $1$ is given as $\bm{z}_1 := (z_{11}, \dots, z_{1G} )$. Suppose observation $i$ is in group $g$, the formulation of factor analysers in the matrix variate case has the density as 
$$\bm{X}_{i} | z_{ig}  = 1  \sim  \mathcal{N}_{r , c} ( \bm{X}_i;  \MM_g, \MU_g + \bm{A}_g\bm{A}_g^{'}, \MV_g + \bm{B}_g\bm{B}^{'}_g).$$
When written in this formulation, the complete data-likelihood is taken to be
$$L( \bm{X}; \bm{\theta} ) =\prod_{i=1}^N \sum_{g=1}^G [\pi_g \varphi ( \bm{X}_i; \MM_g, \MU_g + \bm{A}_g\bm{A}_g^{'}, \MV_g^{'} + \bm{B}_g\bm{B}^{'}_g)]^{z_{ig}}, $$
where $\bm{\theta} = ( \bm{\theta}_1 := (\pi_1, \bm{z}_1, \MM_1, \MU_1,\MV_1, \bm{A}_1, \bm{B}_1),  \dots, \bm{\theta}_G)$. Estimation based on the AECM algorithm is complicated having multiple intermediate steps and algebriac expressions. To maintain clarity, the following is a summarized version where several intermediate steps $\hat{\bm{S}}^A_g \text{ and } \hat{\bm{S}}^B_g$  are ommited. For specifics, see \cite{bifact}. The AECM algorithm consists of three stages. Within the first stage, the complete-data is taken to be the observed matrices $\bm{X}_{1},...,\bm{X}_{N}$, and the component memberships $\bm{z} = (\bm{z}_1,...,\bm{z}_{N})$. The updates for $\pi_g$, and $\MM_g$, for some iteration $t$ are calculated as  
 $$\hat{z}^{(t)}_{ig}= \frac{\pi_g \varphi( \bm{X}_i; \bm{\theta}_g) }{\sum_{h=1}^G \pi_h \varphi( \bm{X}_i; \bm{\theta}_h) },\ \
 \hat{\MM}^{(t)}_g = \frac{ \sum_{i=1}^N \hat{z}^{(t)}_{ig}\bm{X}_i,}{\sum_{i=1}^N \hat{z}^{(t)}_{ig} } \quad \text{and} \quad \hat{\pi}^{(t)}_g = \frac{\sum_{i=1}^N \hat{z}^{(t)}_{ig}}{N}. $$
In the second stage, the complete-data is taken to be the observed $\bm{X}_1, . . . , \bm{X}_N$, the component memberships $\bm{z}$, and the $r \times s$ latent matrices for column factors. In addition, $N^{(t)}_g = \sum_{i=1}^N \hat{z}^{(t)}_{ig} $.  The updates for $\MU$ are taken to be
$$ \hat{\MU}^{(t)}_g = \frac{1}{N^{(t)}_g c} \text{diag} \{ \hat{\bm{S}}^B_g \}.$$
 In the third stage, the complete-data is taken to be the observed $\bm{X}_1, . . . , \bm{X}_N$ , the component memberships $\bm{z}$, and the $c \times v$ latent matrices for column factors allowing to compute estimate for $\MV$ as
$$ \hat{\MV}^{(t)}_g = \frac{1}{N^{(t)}_g r} \text{diag} \{ \hat{\bm{S}}^A_g \} .$$
Convergence of the AECM algorithm is based on the Aitken acceleration criterion \citep{aitken} defined as 
$$ a^{\star (t)} = \frac{l^{(t+1) } - l^{(t)}}{l^{(t)}-l^{(t-1)}}, $$
where $l^{(t)}$ is the observed log likelihood at iteration $t$. Let $ l_\infty^{(t+1)} = l^{(t)} +\frac{l^{(t+1) } - l^{(t)}}{1- a^{\star(t)}} $ be the observed estimate after many iterations at $t+1$. Termination of the algorithm occurs when $ l_\infty^{(t+1)} -  l^{(t)} \in (0,\varepsilon)$ for some pre-specified  $\varepsilon$. Model selection is based on the Bayesian Information Criterion \citep[BIC,][]{schwarz1978}. The BIC is a measure to assess the performance of the model fit, while penalizing for the number of parameters used. For interpretability, the positive scale BIC is used for assessing model performance (larger is better). Let $\rho$ be the number of parameters used. The positive scale BIC is then formulated as  $ \text{BIC}  = 2 l(\bm{\theta}) - \rho \log N.$
With all methods relating to MBI introduced, the clustering problem is formulated as follows. 
Assuming there exists a heterogeneous population of participants of up to $G$ types. Let $\bm{X}_i$ be a force map of participant $i$. Formally, $\mathscr{X} \sim \mathcal{N}^G_{r,c}(\bm{\theta})$  with probability density function 
$$p(\bm{X}_i; \bm{\theta}) = \sum_{g=1}^G \pi_g \varphi( \bm{X}_i; \MM_g, \MU_g + \bm{A}_g\bm{A}_g^{'}, \MV_g^{'} + \bm{B}_g \bm{B}^{'}_g), $$
where $\bm{\theta} =( \bm{\theta}_1 = (\pi_1, \MM_1, \MU_1, \MV_1, \bm{A}_1, \bm{B}_1),  \dots, \bm{\theta}_G) $. 
Estimation of the model is done in accordance with the AECM algorithm as mentioned previously. Once the model has been estimated, classification of participants into one of $G$ types is done in accordance with taking the maximum aposteriori of component memberships. 
The classification $\mathcal{C}_i$ for participant $i$ is given by $\mathcal{C}_i = {\argmax(\hat{z}_{i1}, \dots \hat{z}_{iG}) } $. With each participant classified into a group, we pursue survival analysis to determine differences in lifetime outcomes.

\subsection{Survival Analysis}
As a standard practice, survival analysis in this work is performed by the use of the Cox proportional hazards model \citep{cox1972}.
In addition, from a non-parametric perspective, we also fit Kaplan-Meier (KM) estimators to the data \citep{kaplan1958}. For both models, we test for significance using the log-rank test \citep{kleinbaum2012}. For implementation, see \texttt{survival} and \texttt{survminer} packages \citep{survival-package,survminer-package}. In summary, we use these methods to determine whether or not survival outcomes differ between groups.

\section{Results}

We report our results in chronological order of analysis. We first report our survival analysis on both SW-$0$ and SW-P to establish a baseline understanding of our shift work population. All survival analysis is performed with the R programming language \citep{rlang}. 
 Next, we report our clustering results of SW-P's force maps. All clustering and data manipulation is performed using the Julia programming language \citep{julia}. For an in-depth look on how to interoperate between R and Julia see \cite{mcnicholas19}. Finally, we report our survival analysis results by cluster and sex.

\subsection{Work Shift Survival Analysis}

\begin{figure}[!htb]

  \begin{subfigure}[t]{0.47\textwidth}
        \centering
        \caption{SW-$0$ Kaplan-Meier Curves}
		\includegraphics[scale = 0.45]{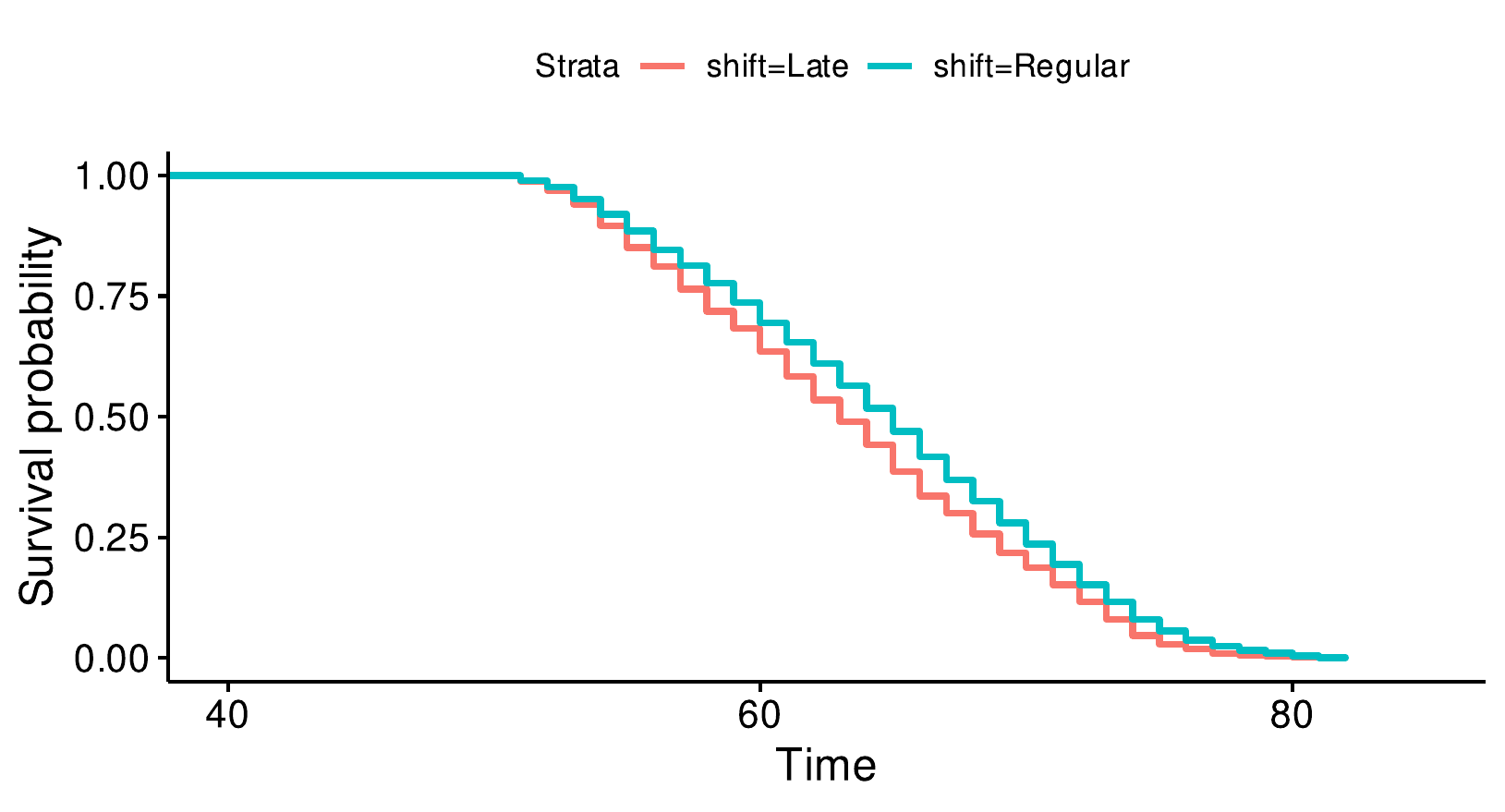}
        \label{fig:sw0_km}
    \end{subfigure} %
      \begin{subfigure}[t]{0.45\textwidth}
        \centering
        \caption{SW-P Kaplan-Meier Curves}
		\includegraphics[scale = 0.45]{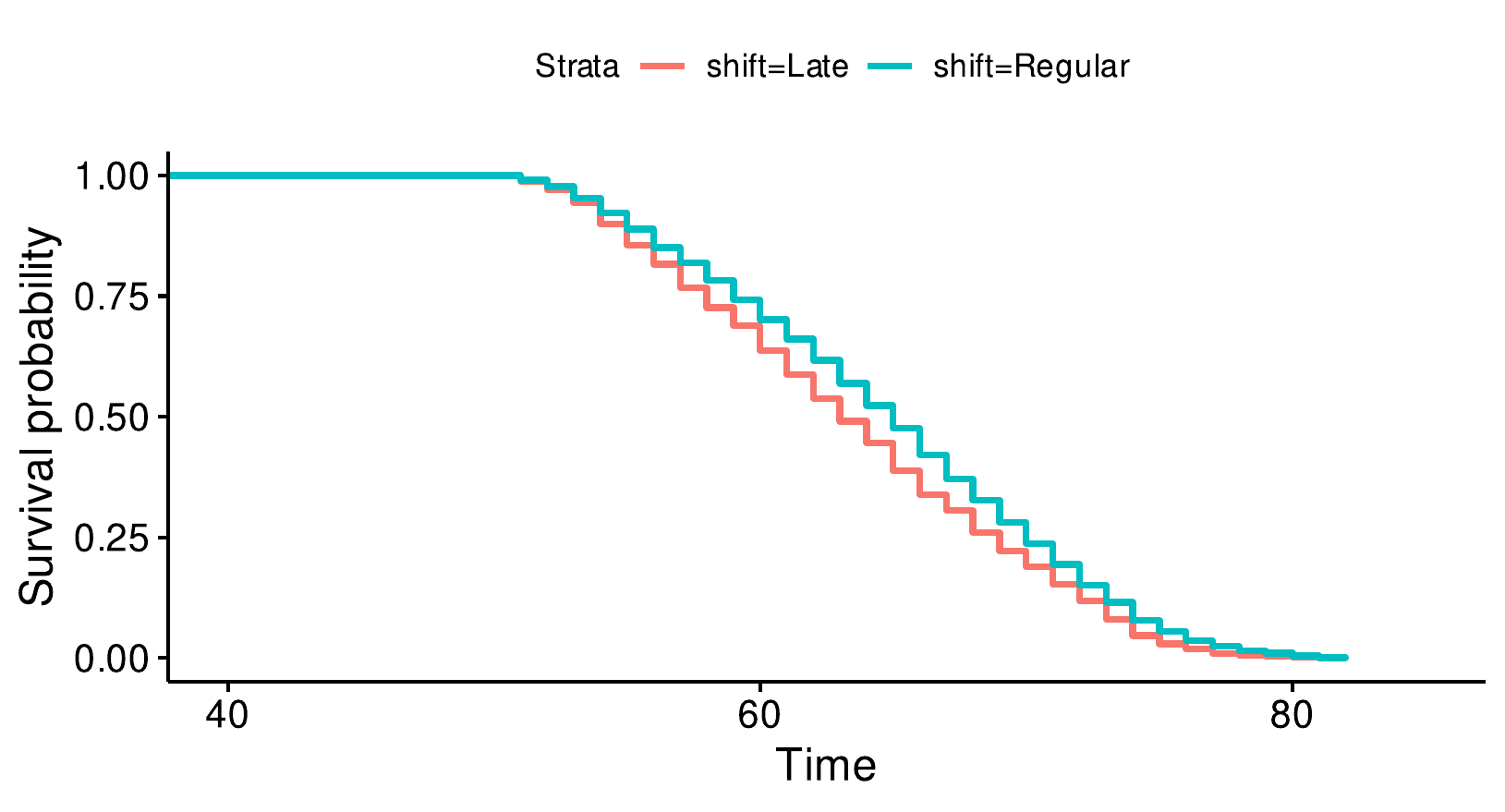}
        \label{fig:swp_km}
    \end{subfigure} %
        ~ \\
               \begin{subfigure}[t]{1.0\textwidth}
        \centering
        \caption{SW-0 Hazard Ratio}
        \hspace*{-2.5cm} \includegraphics[scale = 0.70]{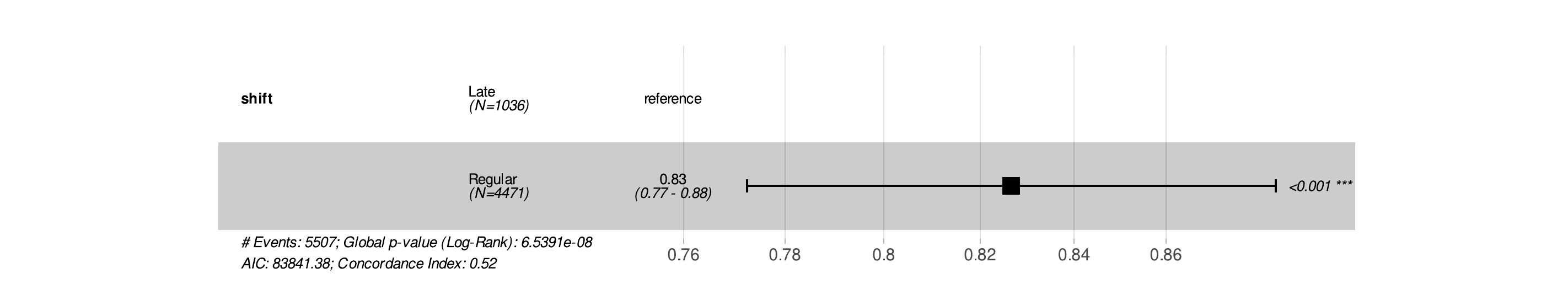}
        \label{fig:swo_cox}
    \end{subfigure}\\
                   \begin{subfigure}[t]{1.0\textwidth}
        \centering
        \caption{SW-P Hazard Ratio}
        \hspace*{-2.5cm} \includegraphics[scale = 0.70]{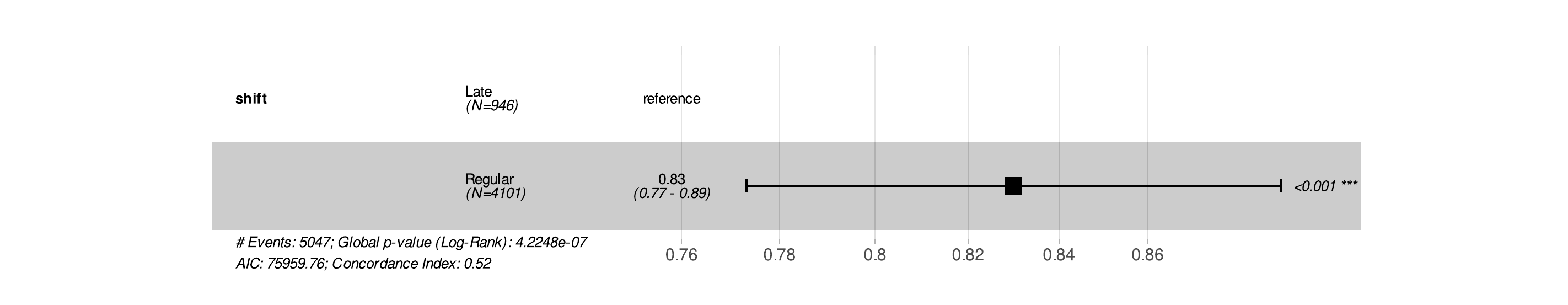}
        \label{fig:swp_cox}
    \end{subfigure}%

\caption{Survival analysis results for populations SW-$0$ and SW-P with time in years.}
\label{fig:survival_analysis_sw}
\end{figure}

Our goal for this portion of our analysis is twofold. First is to reach consensus with the standard occupational health viewpoint that life expectancy suffers when participants defy their natural circadian rhythm. Our two populations, SW-$0$, and SW-P, contain a sufficient number of individuals for analysis to be valid. We fit KM estimators on both SW-$0$ and SW-P. In addition, we performed a log-rank test. Both populations SW-$0$, and SW-P yield statistical significant results indicating the following. Late shift workers have a lower survival rate than regular shift workers. Figures \ref{fig:sw0_km}, and \ref{fig:swp_km} illustrate the KM estimates over time, where time is measured in years. The curves between both of the populations are virtually the same. In both plots, both types of shift workers have the same survival estimate until the age of $57$. At this point, the two shift worker groups start to diverge, until reaching the age of $80$. The log-rank tests for both SW-$0$, and SW-P yield a p-value of less than $0.001$. Despite there being a loss in the number of participants between SW-$0$ and SW-P (see section \ref{sec:2.2}), SW-P still retains the relevant survival outcomes of its predecessor SW-$0$.

The results of the Cox regression yields similar results. Figures \ref{fig:swo_cox}, and \ref{fig:swp_cox} report the hazard ratios in the SW-$0$, and SW-P populations respectively. Again we find that the regular shift workers have a better survival outcome than late shift workers. As expected, there are ostensibly no differences between the results of SW-0, and SW-P. However, the range of SW-P is slightly higher with a right bound of $0.89$, than SW-0's $0.88$. In summary, both analyses yield statistically significant results, and conclude the same survival outcomes. Regular shift workers have a hazard rate that is less than late shift workers. This is in-line with the general consensus of \cite{gu2015}, and \cite{ijaz2013}.

\subsection{Force Map Clusters} \label{sec22}
As a preparation step, we perform a logit transformation on each entry across all force maps to map the domain from $\mathbb{R}[0,1]$ to $\mathbb{R}$. This allows the MBI model to have a better fit for the data while retaining the force map structure across each participant. Another preparation step is regarding initialization of cluster memberships. There are a number of initializations possible, however for simplicity we use $k$-means, and random soft ($k$ meaning the number of groups in this context). For specifics regarding these two methods, see \cite{mcnich15}. 
There are a number of settings to consider for the MBI model. We search for the best model with the following settings: $q = \{1,\dots,7\},s = \{1,\dots,7\}, G = \{1, \dots , 6\}$. Table \ref{table:bic-compare} shows the results of our model search. According to BIC, we report the top 5 performing MBI models. The best model is selected to be $G=3$ for the number of groups, a BIC of $-9527009$, and $q=5,s=4$ for the dimension of latent factors.

 \begin{table}
\parbox{.45\linewidth}{
\centering
\caption{Top 5 performing MBI models.}
\label{table:bic-compare}
\begin{tabular}{|c|ccc|}
\hline
BIC       & $G$ & $q$ & $s$ \\
\hline
$\bm{-9527009} $  & $\bm{3}$ & $\bm{5}$ & $\bm{4}$ \\
$-9753041$  & $4$ & $4$ & $4$ \\
$-9928201$  & $4$ & $5$ & $3$ \\
$-10075763$ & $3$ & $5$ & $3$ \\
$-10188673 $& $3$ & $4$ & $4$ \\
\hline
\end{tabular}
}
\hfill
\parbox{.45\linewidth}{
\centering
\caption{Clustering Results.}
\label{table:cluster_descript}
\begin{tabular}{|c|c|c|c|}
\hline
Cluster & SW-1 & SW-2 & SW-3 \\ 
\hline
$\pi_g$ & $0.0806$ & $0.3385$ & $0.5809$ \\
$n_g$ & $407$ & $1,708$ & $2,911$ \\ 
\hline
\end{tabular}
}
\end{table}

\begin{figure}[!htb]
\centering
\hspace*{-1cm}
  \begin{subfigure}[t]{0.33\textwidth}
        \centering
        \caption{SW-1}
		\includegraphics[scale = 0.32]{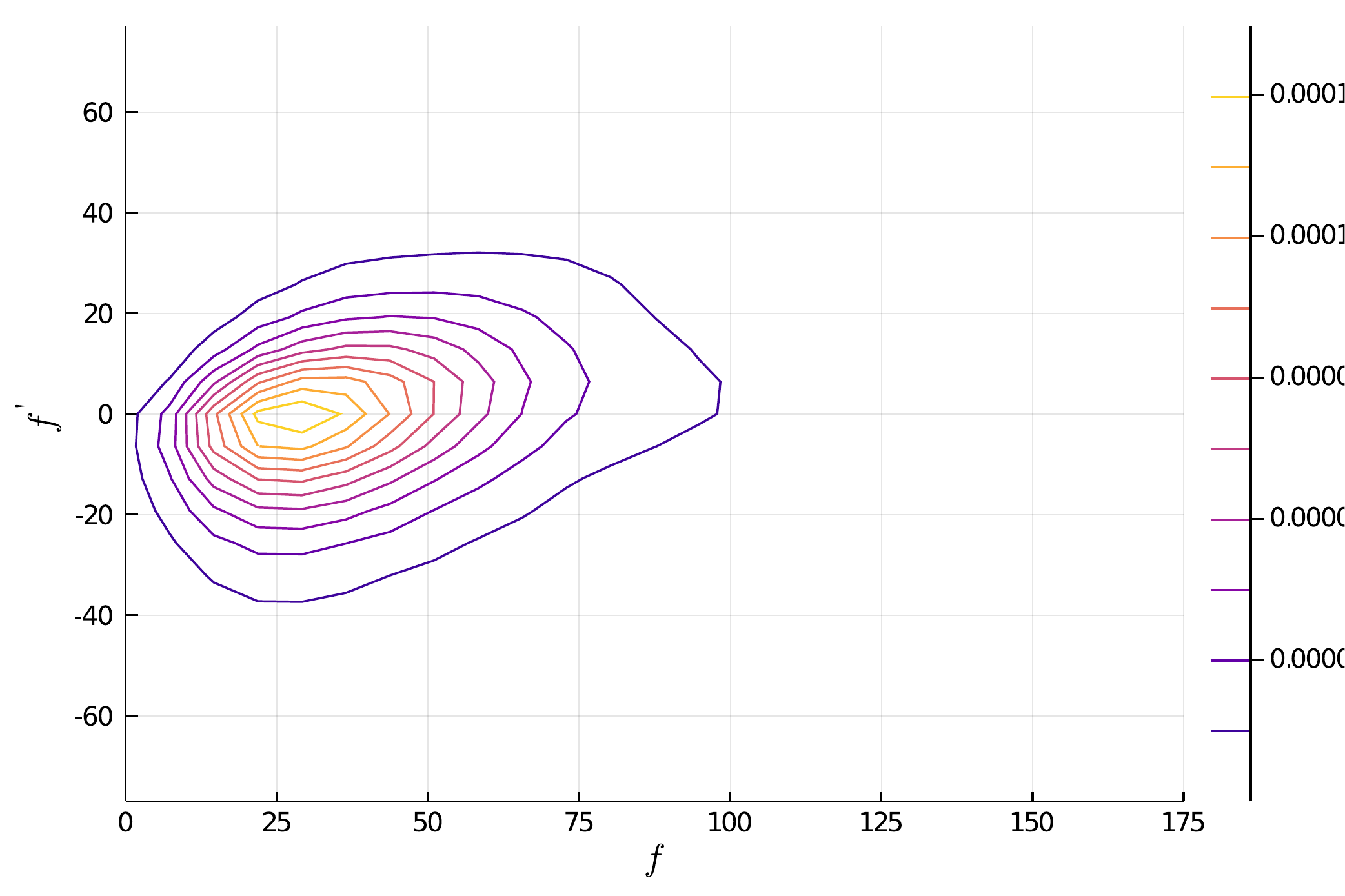}
        \label{fig:cluster-1}
    \end{subfigure} %
        ~ 
        \begin{subfigure}[t]{0.32\textwidth}
        \centering
        \caption{SW-2}
		\includegraphics[scale = 0.32]{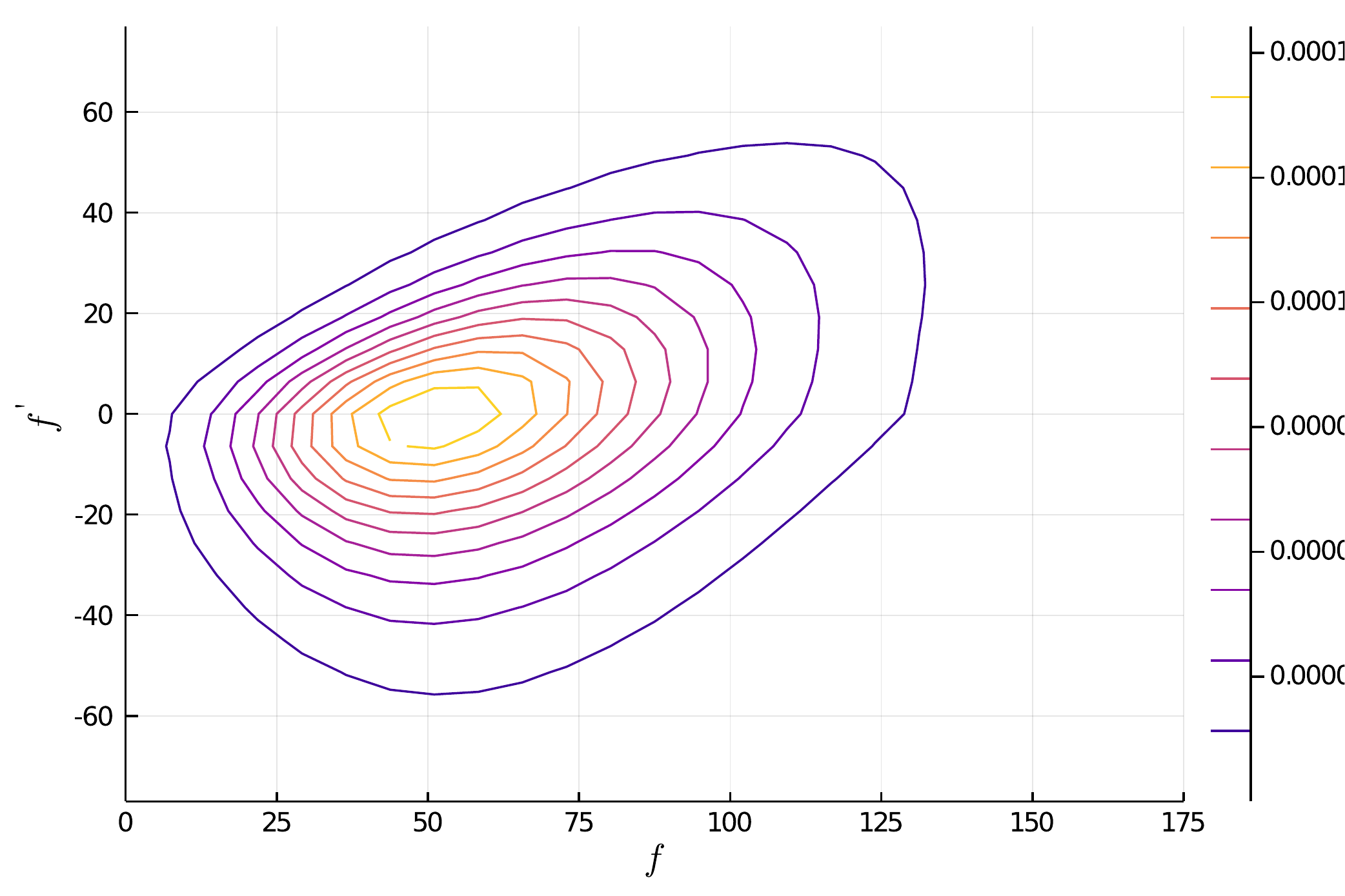}
        \label{fig:cluster-2}
    \end{subfigure}%
           \begin{subfigure}[t]{0.32\textwidth}
        \centering
        \caption{SW-3}
		\includegraphics[scale = 0.32]{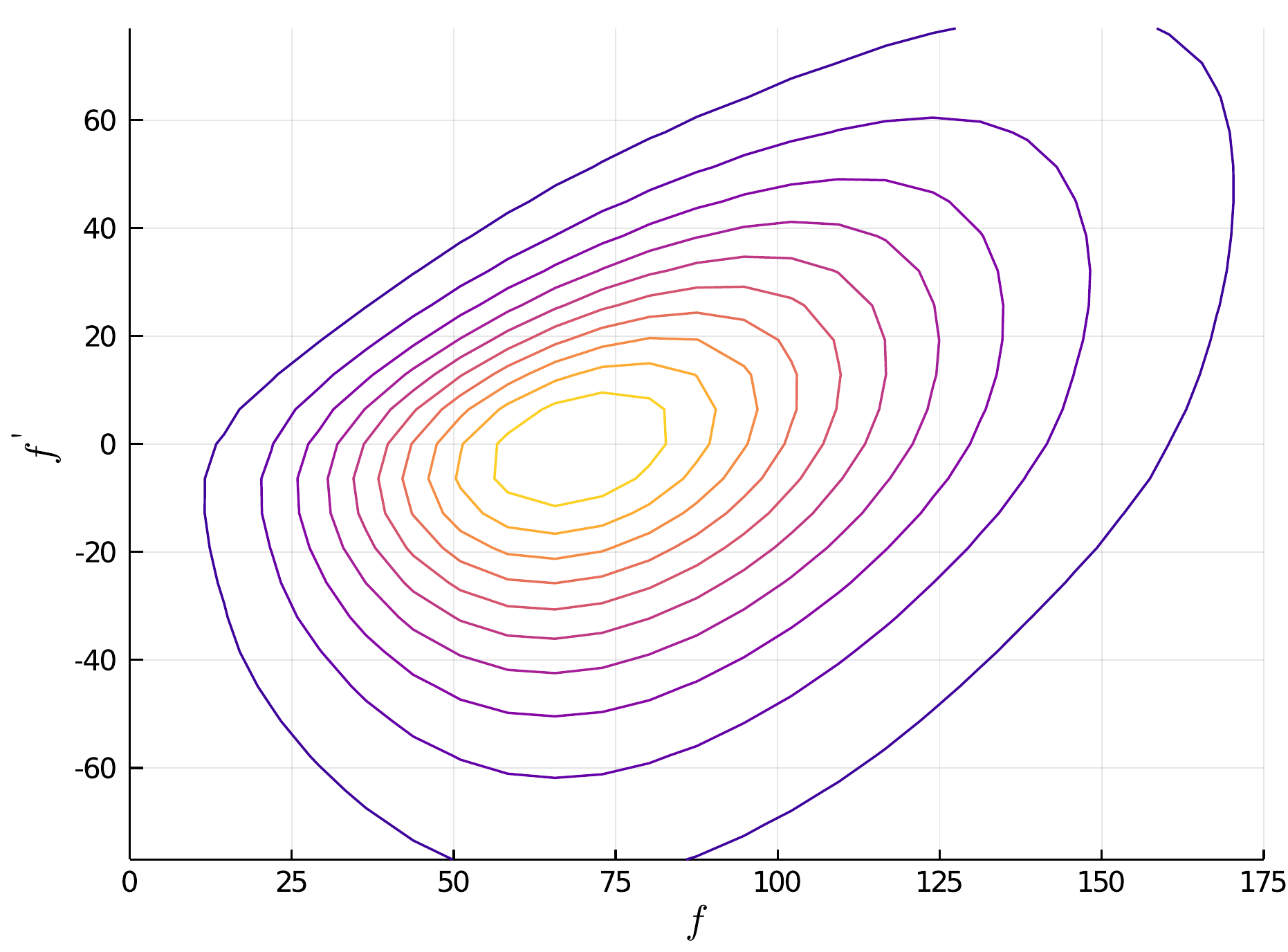}
        \label{fig:cluster-3}
    \end{subfigure}%

\caption{Mean force map $\MM_g$ contours for moderate task behaviour, by cluster $g$.}
\label{fig:forcemaps-cluster}
\end{figure}

By clustering each participant of SW-P into one of three groups, we define three new populations of shift workers as SW-$1$, SW-$2$, and SW-$3$. Table \ref{table:cluster_descript} displays the cluster results for each group. There are $407$ participants clustered in SW-$1$ which account for smallest portion of the total population of SW-P. In contrast, there are $1,708$ participants clustered into SW-$2$. These participants account for $33.85\%$ of SW-P. Finally, SW-$3$ accounts for the majority with $2,911$ participants. We draw attention to the mean force map of each cluster in Figure \ref{fig:forcemaps-cluster}. The mean force map for SW-$1$ seen in Figure \ref{fig:cluster-1}, has most of the readings around $25$ force units and is quite concentrated. In contrast, the mean force map for SW-$2$ has most of the readings centred around $50$ force units but is quite spread. Finally, the mean force map of SW-$3$ has most of the readings centred around $73$ force units, and is the most spread by far. The interpretation of each force map is as follows. Participants who exhibit stronger physical activity during moderate tasks have higher force readings. As a result, their force map becomes variable and less concentrated. Each cluster from SW-$1$ to SW-$3$ embodies a gradual increase in physical exertions. In SW-$1$ we have participants who have the lowest level of physical exertion, while in SW-$3$ we have the highest.

\newpage
In summary, our clusters embody the heterogeneous levels of physical behaviour within the SW-P population. Now that we have established our populations of interest,
we further investigate our clustering results using survival analysis.

\subsection{Cluster Survival Analysis} \label{section:cluster}

\begin{figure}[!htb]
\centering
\hspace*{-1cm} \includegraphics[scale=0.65]{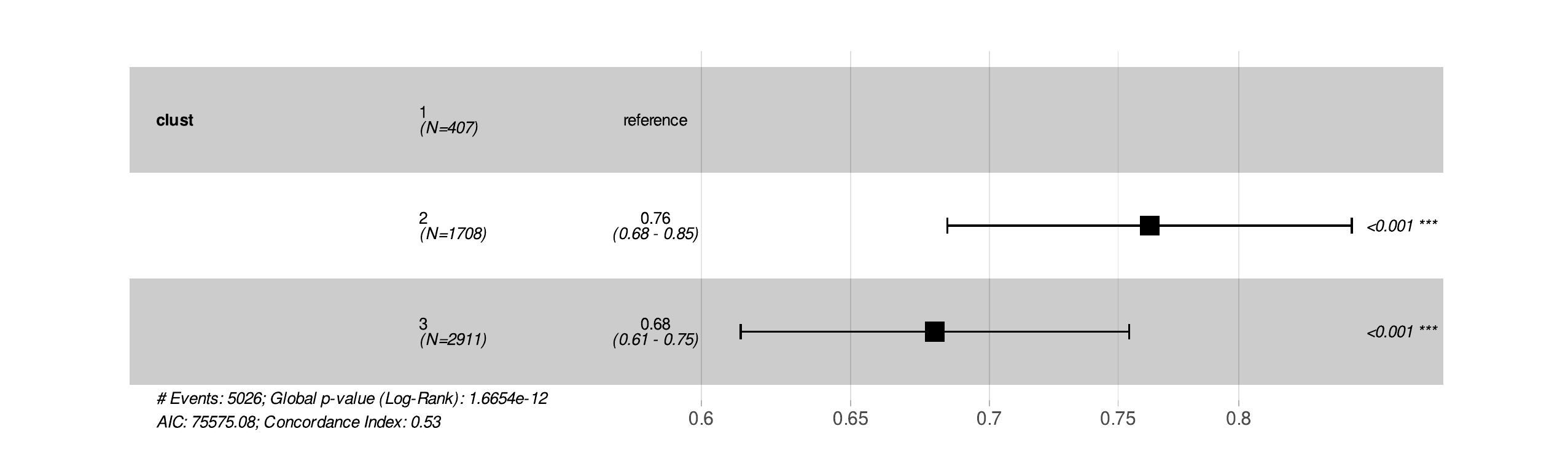}
\caption{Cox regression results by cluster.}
\label{fig:cox_by_cluster}
\end{figure}

Using our previously defined populations SW-1 to SW-3, we proceed with survival analysis by cluster. Figure \ref{fig:cox_by_cluster} shows three decreasing levels of risk. We report a statistically significant global log-rank a $p$-value of less than $0.001$. Cluster $1$ is used as a reference for the cox proportional hazards model. Both cluster $2$ and cluster $3$ have a lower risk than the reference cluster.

We further fit a Cox regression between cluster $2$ and $3$, this time using cluster $2$ as a reference. Here, we see that cluster $3$ has indeed a lower risk than cluster $2$. Again, the log-rank test shows a statistically significant result.

\begin{figure}[!htb]
\centering
z\hspace*{-3cm} \includegraphics[scale=0.60]{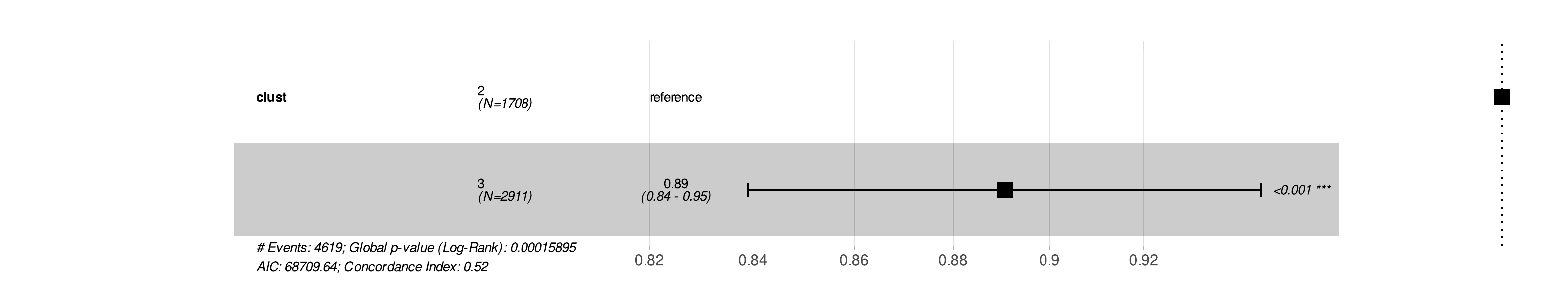}
\caption{Cox regression results between SW-$2$ and SW-$3$.}
\label{fig:cox_by_cluster23}
\end{figure}
\begin{figure}[!htb]
\centering
\hspace*{-1.75cm} \includegraphics[scale=0.60]{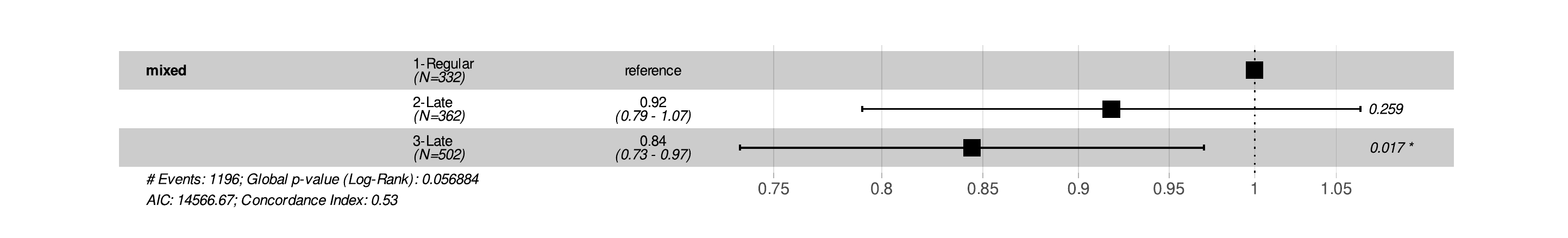}
\caption{Cox regression results for mixed populations.}
\label{fig:cox_mix}
\end{figure}

\newpage
To summarize, we describe three decreasing levels of risk where each cluster has a statistically significant difference in hazard rates. These survival outcomes are consistent with our force map analysis in Section \ref{sec22}. Cluster $1$ force maps have the lowest level of physical exertion which increases between clusters. Similarly in our survival analysis, cluster $1$ participants have the highest risk that decreases from cluster $2$ to $3$. We also fit a cox regression to analyse a mixed interaction between clusters and type of shift work. Particularly to individuals belonging to cluster 1, and engaging in regular work, as well as individuals who belong to cluster 2 and 3, that engage in late shift work. This enables us investigate if individuals 
who engage in late shift work and have higher physical activity, may offset their risk when compared to regular shift workers that do not engage in physical activity. Figure 
\ref{fig:cox_mix} shows the hazard ratios of these populations.

\begin{figure}[!htb]
\centering
\hspace*{-1.75cm} \includegraphics[scale=0.60]{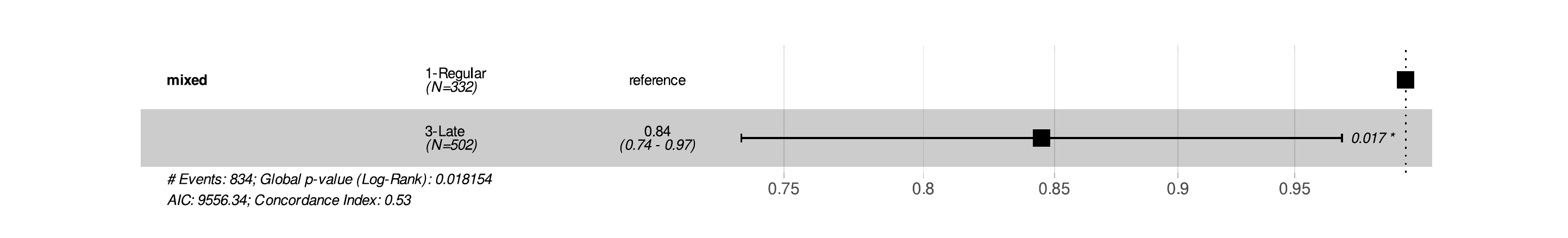}
\caption{Cox regression results for mixed populations between regular and late shift work.}
\label{fig:cox_mix13}
\end{figure}

\begin{figure}[!htb]
\centering
\includegraphics[scale=0.57]{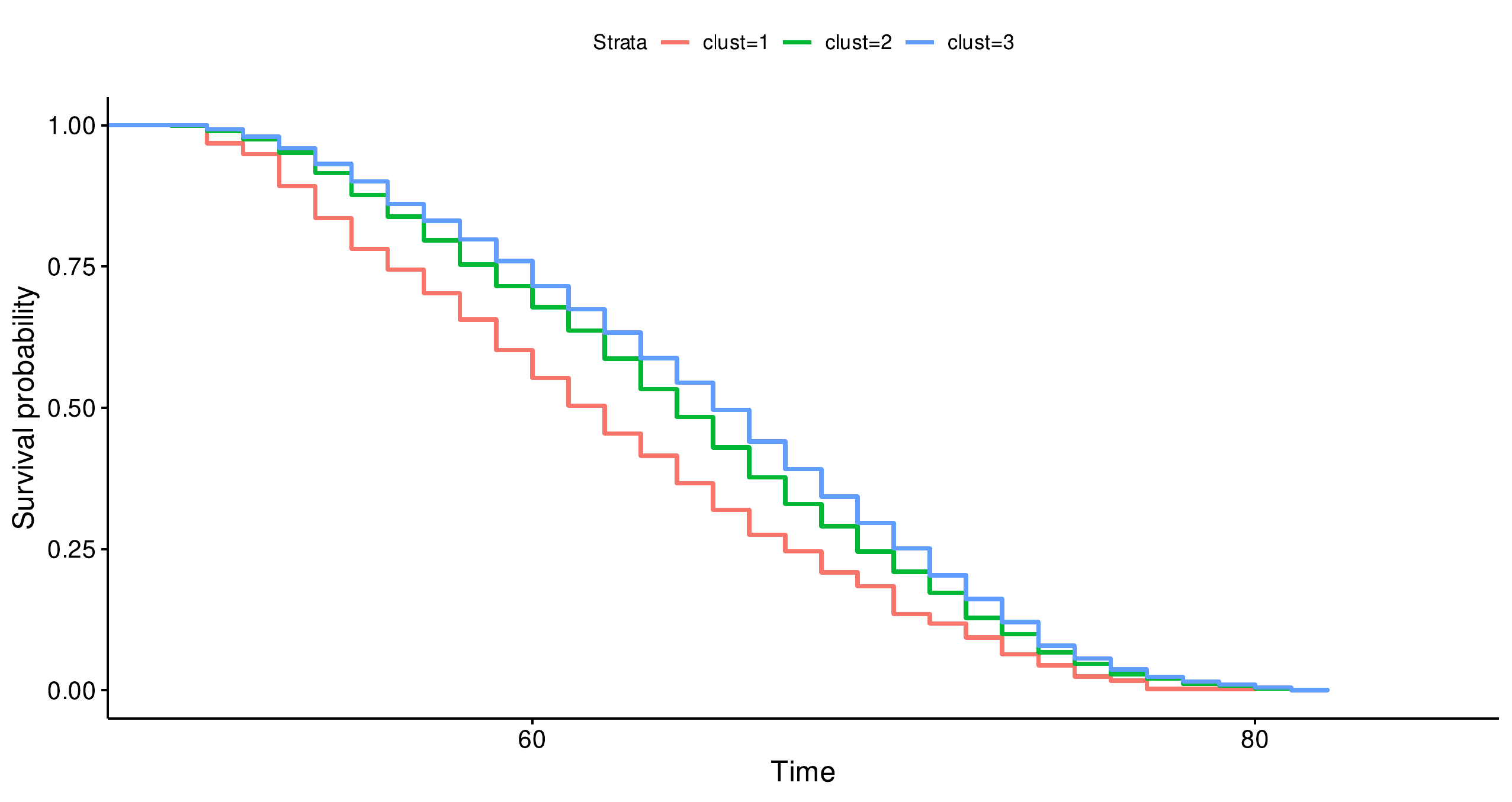}
\caption{Kaplan-Meier Curves for each cluster population with time in years.}
\label{fig:swc_km}
\end{figure}

A statistically significant difference between cluster 1 regular shift workers, and cluster 3 late shift workers is reported. We further fit the same cox regression only on these two populations. As expected, Figure \ref{fig:cox_mix13} shows a statistically significant difference between rates of these populations. Consistently, the KM estimators capture the same decreasing survival outcome as shown in Figure \ref{fig:swc_km}.

\begin{figure}[!htb]
\centering
\includegraphics[scale=0.50]{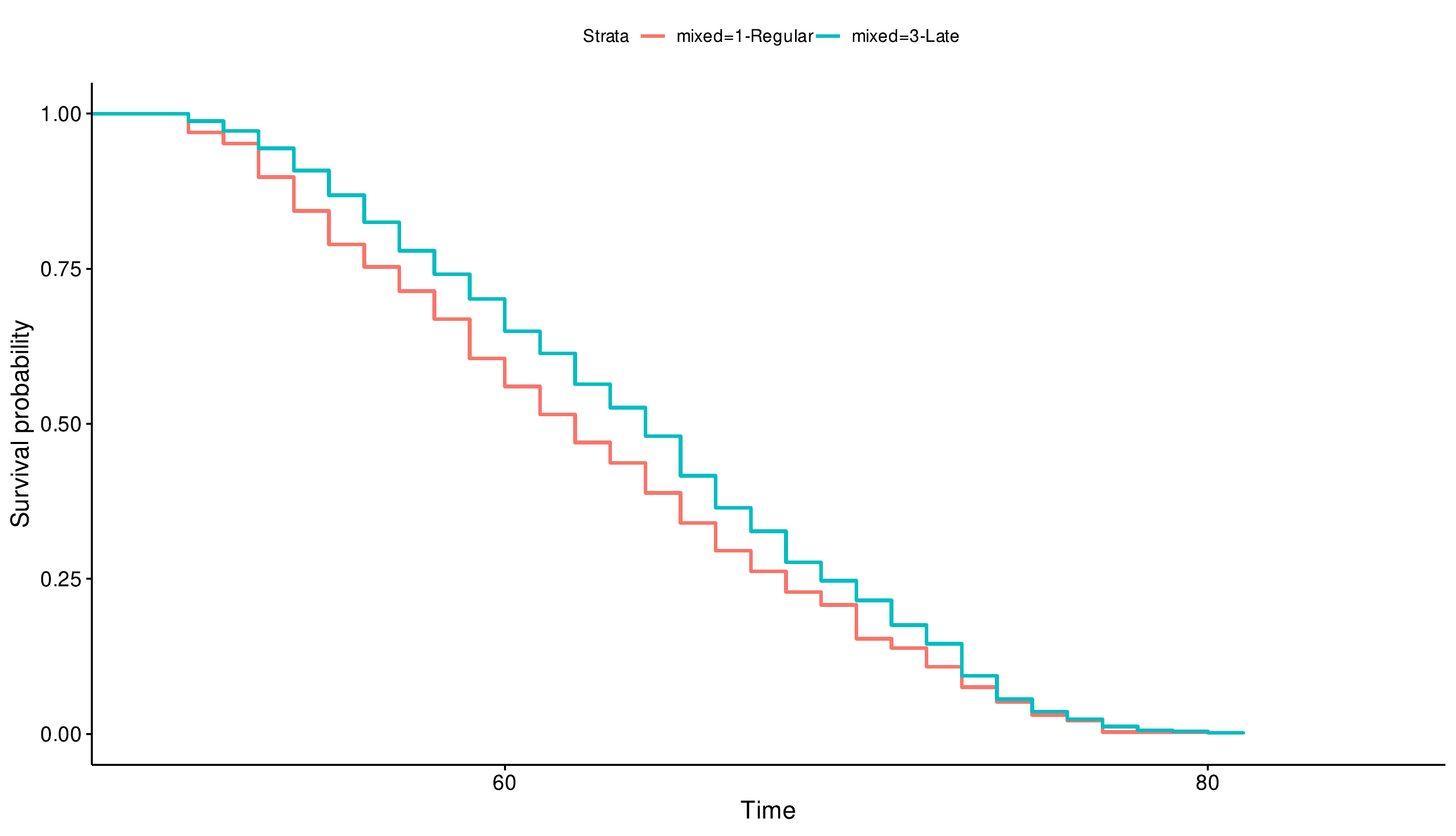}
\caption{Kaplan-Meier Curves for mixed cluster populations with time in years.}
\label{fig:swc_km13}
\end{figure}

We interpret these results as follows. Individuals who are defined as late shift workers, and exert the highest level of physical activity, have a better survival outcome than individuals who undergo regular shift work, and exert the lowest physical activity. Figure \ref{fig:swc_km13} shows separated KM curves in consensus with our cox regression results in Figure \ref{fig:cox_mix13}. 
 Generally, across all KM curves, there is no evidence of censoring within each population. Each cluster or sub-group contains an ample amount of individuals to reliably estimate survival probabilities using a KM estimator.

\newpage
\subsection{Cluster Survival Analysis by Sex}
\begin{figure}[!htb]
\centering
\includegraphics[scale=0.55]{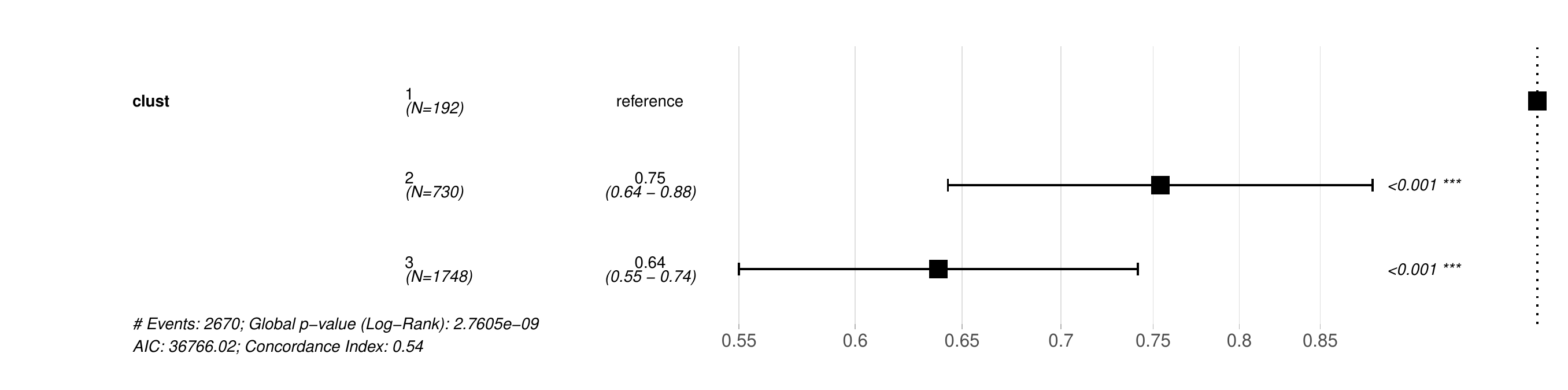}
\caption{Hazard ratios for females by cluster.}
\label{fig:sex_female_hazard}
\end{figure}

\begin{figure}[!htb]
\centering
\includegraphics[scale=0.55]{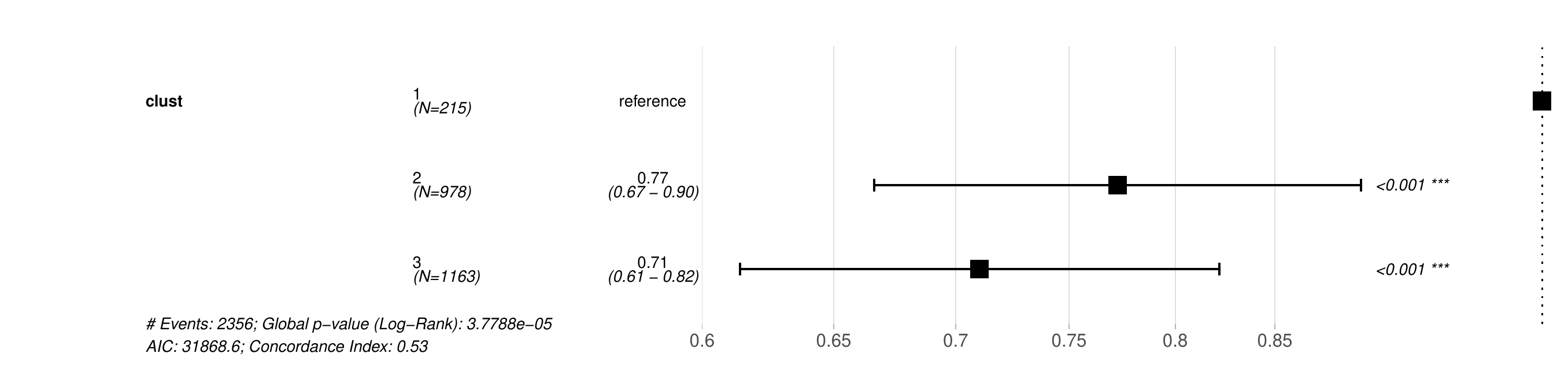}
\caption{Hazard ratios for males by cluster.}
\label{fig:sex_male_hazard}
\end{figure}

When taking into consideration sex differences, consider the proportion of males and females within each cluster. Table \ref{table:sex-clust} shows a clear majority of males among clusters 1 and 2, while a female majority among cluster 3. As mentioned previously, there are more favourable survival outcomes for females when compared to males. However, when performing Cox regression by sex, we have the same three decreasing levels of risk as in Section \ref{section:cluster}.  Figures \ref{fig:sex_female_hazard} and \ref{fig:sex_male_hazard} show statistically significant differences between each of the clusters.  However, females have stronger differences in their hazard ratios when compared to males.

\begin{figure}[!htb]
\centering
\includegraphics[scale=0.55]{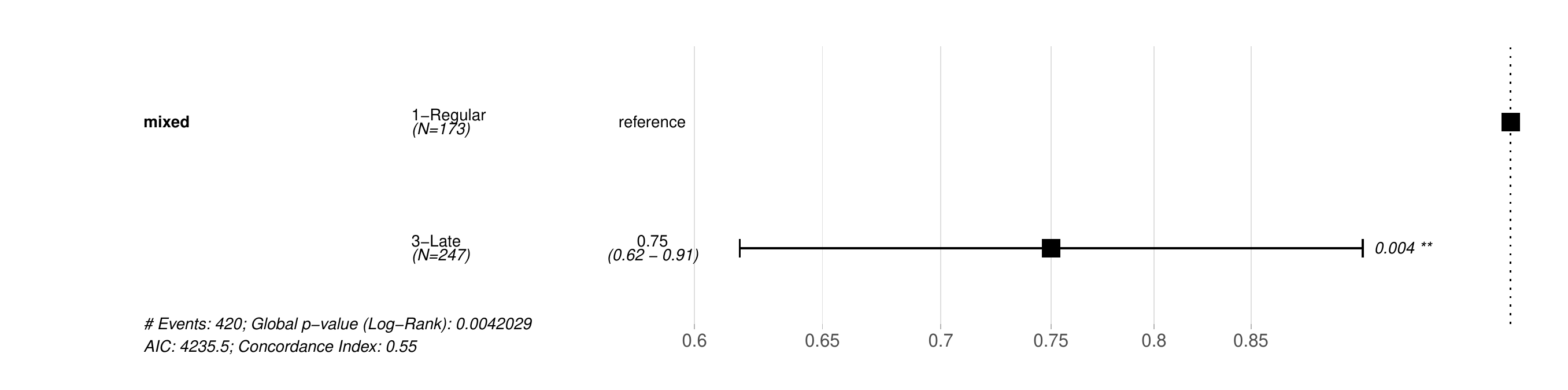}
\caption{Hazard ratios for females by mixed groups between type of shift work, and cluster.}
\label{fig:sex_female_hazard_2}
\end{figure}

\begin{figure}[!htb]
\centering
\includegraphics[scale=0.55]{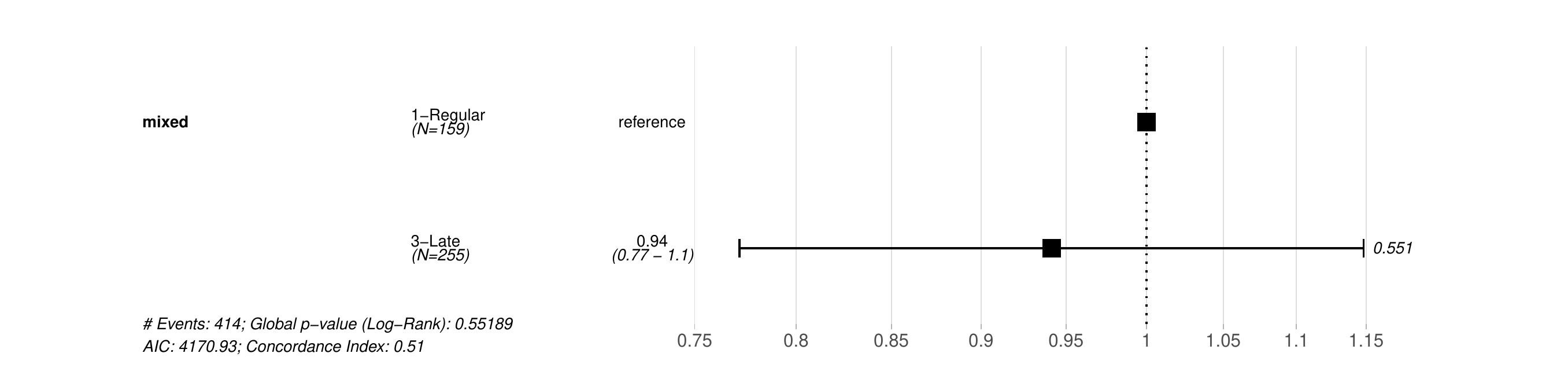}
\caption{Hazard ratios for males by mixed groups between type of shift work, and cluster.}
\label{fig:sex_male_hazard_2}
\end{figure}

When attempting to reproduce the protective effect of physical activity in Section \ref{section:cluster}, we discover an interesting result.  Consider Figures \ref{fig:sex_female_hazard_2}, and \ref{fig:sex_male_hazard_2}. For females, there is a clear benefit for physical activity on survival outcomes. We report an offset in risk by a factor of $0.75$.  Unfortunately, this does not translate for males, as their hazard ratios are deemed to be statistically insignificant.

\begin{table}[!htb]
\centering
\caption{Participant sex counts by cluster.}
\label{table:sex-clust}
\begin{tabular}{|c|c|c|}
\hline
Cluster &  Female &  Male \\
\hline
1 &   $192\hspace{3pt} (47.17\%)$  & $ 215 \hspace{3pt} ( 52.83 \% )$ \\
2 &  $730 \hspace{3pt} (42.74 \%)$  & $978 \hspace{3pt} ( 57.26 \%)$ \\
3 &  $1,748 \hspace{3pt} (60.04 \%)$  & $39.96  \hspace{3pt} ( 43.96 \%)$ \\
\hline
\end{tabular}

\end{table}

\section{Conclusion}
For individuals that defy their circadian rhythm, through for example, working shift patterns, our results indicate that an increase in physical activity may be enough to offset the additional risk associated with late shift work. Although our results indicate that there are no statistically significant differences for males, it beckons the question of why this protective effect is only present for SW-P females. During our clustering procedure, male and female force maps were not separated, and therefore, the stronger signal from the female population may have caused this phenomenon. Nevertheless, a segmentation by sex approach can be easily incorporated into our methodology.

By considering the sheer size and computational cost of managing this dataset, our approach is feasible and scalable for larger cohorts. The methodology proposes a novel set of statistical tools for capturing participant behaviour within accelerometer data. The force maps characterize participant behaviour and allow for visually interpretable results. Furthermore, by introducing this matrix variate object, we standardize an otherwise incomparable time series; allowing for comparisons to be made between participants. 

In conclusion, we utilize accelerometer data in a manageable and efficient process to cluster participants accordingly. The results of these clusters indicate valid survival outcomes with great potential for improving health from an occupational perspective. We developed a low-dimensional feature that is interpretable, and captures participant behaviour in a unique way. Future work, from a clinical perspective, may consider the possibility of investigating the protective nature of physical activity for those who defy their circadian rhythm.

{\small

\section*{Acknowledgements}
This work was supported by Ontario Canada Graduate Scholarship (Po\v cu\v ca), the Canada Research Chairs program (McNicholas), and an E.W.R~Steacie Memorial Fellowship.

\bibliographystyle{chicago}
\bibliography{tmn_paper}}

\end{document}